\documentclass[12pt]{article}
\usepackage[pdftex]{pict2e}

\usepackage{mathrsfs,amsthm,amsmath,latexsym,amssymb,amsfonts,amscd,graphicx,latexsym}
\usepackage{graphics,lscape,fancyhdr,array,stmaryrd,euscript}
\usepackage{ifthen,empheq,slashed}
\usepackage{color}
\numberwithin{equation}{section}
\usepackage{setspace}
\usepackage{cite}

\DeclareFontFamily{U}{mathx}{\hyphenchar\font45 }
\DeclareFontShape{U}{mathx}{m}{n}{
	<-> mathx5
}{}
\DeclareSymbolFont{mathx}{U}{mathx}{m}{n}
\DeclareMathAccent{\wideTilde}{0}{mathx}{"72}

\usepackage[hang,flushmargin]{footmisc} 
\usepackage[citecolor=purple,linkcolor=blue,urlcolor=blue,colorlinks=true,filecolor=green,
pdfstartview=FitV, 
bookmarksopen=true,
hyperfootnotes=true]{hyperref}

\usepackage{makecell}
\usepackage{geometry}

\newgeometry{left=.8in,right=.8in,top=.8in,bottom=.85in}
\usepackage[most]{tcolorbox}
\usepackage{enumitem}
\usepackage{footnotebackref}

\usepackage{soul}

\usepackage{tikz}
\usepackage{xcolor}
\usepackage{colortbl}
\usepackage{pgfplots}
\pgfplotsset{compat=1.16}

\definecolor{galilei}{RGB}{255,255,153}
\definecolor{minkowski}{RGB}{204,255,204}
\definecolor{carroll}{RGB}{255,204,204}

\usetikzlibrary{calc}
\usetikzlibrary{arrows.meta, positioning}

\newcommand\boxedB[1]{{\setlength\fboxsep{5pt}\boxed{#1}}}

\newcommand{\xdownarrow}[1]{%
	{\left\Downarrow\vbox to #1{}\right.\kern-\nulldelimiterspace}
}

\usepackage{multirow}
\usepackage{array}
\usepackage{makecell}

\renewcommand{\arraystretch}{3.3} 

\definecolor{yell}{cmyk}{0,0,0,1}

\usepackage{lipsum} 
\usepackage{abstract}

\def\be {\begin{equation}}
	\def\ee {\end{equation}}
\def\le {\left}
\def\ri {\right}
\def\p {\partial}
\def\d {\delta}
\def\m {\mu}
\def\n {\nu}
\def\s {\sigma}
\def\a {\alpha}
\def\b {\beta}
\def\r {\rho}
 
\def\ga {\gamma}
\def\b {\beta}

\def\pc {_{_\mathrm{PC}}}
\def\mb{\mathbf}
\def\1{_{_1}}
\def\2{_{_2}}

\def\gt {\gamma_{_{_t}}}
\def\N {_{\!_N}}
\def\nn {_{_{N}}}

\begin{document}


\begin{center}
	
	{\bfseries \Large{Post-Carrollian Mechanics, Ideal Gas, and Gravity}} \vspace{-8pt}
	\par\noindent\rule{440pt}{2pt}\\
	
	\vskip 0.04\textheight

	Mojtaba \textsc{Najafizadeh}{}{\,$^{1,\,2}$}

	\vskip 0.01\textheight

	\vspace*{5pt}
	${}^{1}${\small\em 
		School of Physics, Institute for Research in Fundamental Sciences (IPM), \\ 
		P.O.Box 19395-5531, Tehran, Iran}
	
	\vskip 0.01\textheight
	
	\vspace*{5pt}
	${}^{2}${\small\em Department of Physics, Faculty of Science, Ferdowsi University of Mashhad,\\ 
		P.O.Box 1436, Mashhad, Iran}
	
	
	\vskip 0.01\textheight
	
	\href{mailto:mnajafizadeh@gmail.com}{mnajafizadeh@ipm.ir (@gmail.com)}
	
	\vskip 0.05\textheight

	{\bf Abstract }
	
\end{center}
\setlength{\leftskip}{0cm} 
\setlength{\rightskip}{0cm} 

\noindent {\small Energy and momentum in Newtonian mechanics have the familiar relations, ($\mathrm{E}=mv^2/2$) and ($\mathbf{P}=m\mathbf{v}$), derived from the non-relativistic limit of special relativity. In this study, we find the corresponding relations to formulate the so-called ``post-Carrollian mechanics'' by applying the ultra-relativistic limit to tachyon theory, resulting in ($\mathrm{E}={m\,c^{\,3}}/{v}$) and ($\mathbf{P}=\mathbf{\hat{v}}\,{m\,c^{\,3}}/{2\,v^{\,2}}$). Using these, we determine the energy-momentum relation and investigate the thermodynamics of an ideal gas composed of post-Carroll particles. Moreover, by applying the ultra-relativistic limit to Einstein's equations coupled to tachyon dust, we find the post-Carrollian gravitational potential. Finally, utilizing the geodesic equation, we determine the post-Carrollian gravitational field, which unlike the Newtonian case is found to be radially outward.}

\vskip 0.03\textheight
\noindent{\small\textsc{Keywords}: {Post-Carroll particles, Post-Carrollian mechanics, Post-Carrollian Ideal gas, Post-Carrollian Gravitational field}}

\newpage
\small\tableofcontents
\newpage

\section{Introduction}

The study of Carroll\footnote{The name ``Carroll'' is inspired by the author of ``{\it Alice's Adventures in Wonderland\,}'', Lewis Carroll.} algebra \cite{Levy1965,SenGupta:1966qer,Bacry:1968zf} (see also \cite{Figueroa-OFarrill:2017ycu,Figueroa-OFarrill:2017tcy}) and its potential applications is motivated by several factors. One notable motivation is the discovery that the Carrollian conformal algebra is isomorphic to the Bondi-Metzner-Sachs (BMS) algebra in one dimension higher \cite{Duval:2014uoa, Duval:2014uva, Duval:2014lpa}. This prompted a surge of interest for this intriguing structure from other perspectives with different motivations. For instance, see some recent works \cite{Bagchi:2010zz,Bergshoeff:2014jla,Duval:2017els,deBoer:2017ing,Donnay:2019jiz,deBoer:2021jej,Henneaux:2021yzg,Bergshoeff:2022qkx,Koutrolikos:2023evq,Bagchi:2022eui,Baiguera:2022lsw,Marsot:2022imf,deBoer:2023fnj,Ecker:2023uwm,Ciambelli:2023tzb,Figueroa-OFarrill:2023vbj,Figueroa-OFarrill:2023qty,Bergshoeff:2023vfd,Aggarwal:2024gfb,Ecker:2024czx,Banerjee:2024jub,Harksen:2024bnh,Casalbuoni:2024jmj,Najafizadeh:2024imn,Afshar:2024llh,Ekiz:2025hdn,Bagchi:2025vri} 
and references therein, particularly those related to Carroll gravity \cite{Henneaux:1979vn,Hartong:2015xda,Bergshoeff:2017btm,Hansen:2021fxi,Guerrieri:2021cdz,Campoleoni:2022ebj,Figueroa-OFarrill:2022mcy,Sengupta:2022rbd,March:2024zck,Ecker:2024czh,Ecker:2025ncp}.

Based on the literature, there are two distinct types of Carroll particles: electric and magnetic (also known as time-like and space-like respectively). For a review, see e.g. \cite{Henneaux:2021yzg,deBoer:2021jej,Bergshoeff:2022qkx,Koutrolikos:2023evq} and references therein. Electric Carroll particles have zero velocity, making them localized and fixed in space, which is not the subject of discussion in this work. On the other hand, magnetic Carroll particles are those with non-zero velocity, with $v>c$, and hence they are tachyonic \cite{deBoer:2021jej}. The energy and momentum of these particles are given by $\mathrm{E}=0$ and $\mathbf{P}=mc\,\mathbf{\hat{v}}$ \cite{deBoer:2021jej}\footnote{The momentum is indeed given by $\mathbf{P}=\pm\,mc\,\mathbf{\hat{v}}$ \cite{deBoer:2021jej}, however, we adopt the positive sign for simplicity.}. As we shall demonstrate in the following section, the energy and momentum expansion of tachyons in powers of $c$ can be expressed as
\begin{align}
	\mathrm{E}~&=\qquad 0 \qquad\,\,~~~+~~~~~\quad\frac{mc^3}{v}~~~~~~~~~+~~~\mathcal{O}(c^5)\,,\label{tee}\\[8pt]
	\mathbf{P}~&=~~~~mc\,\mathbf{\hat{v}}~~~~~~~+~~~\quad\frac{1}{2}\,\frac{m\,c^{\,3}}{v^{\,2}}\,\mathbf{\hat{v}}~~~~~+~~~\mathcal{O}(c^5)\,.\label{tm}\\[-5pt]
	&~~\underbrace{~~~\quad\qquad}_{\substack{\\[1pt]\text{leading terms $\equiv$}\\[2pt] \text{magnetic Carroll}}}~~\quad \underbrace{~~~\qquad\qquad}_{\substack{\\[1pt]\text{subleading terms $\equiv$}\\[2pt] \text{post-magnetic Carroll}}}\nonumber
\end{align}
These expansions illustrate that the leading terms correspond to the energy and momentum of magnetic Carroll particles, emerging from the $\frac{c}{v} \to 0$ limit (Carroll limit). However, our focus in this work is on the subleading terms, including corrections to the magnetic Carroll theory, which we refer to as ``post-magnetic Carroll'' contributions, where $\frac{c}{v} \neq 0$. For clarity and simplicity, we will henceforth omit the term ``magnetic'' and refer to these subleading contributions simply as ``post-Carroll''—a term first introduced in \cite{Ecker:2025ncp}, where corrections to the magnetic Carroll theory are considered.

\begin{figure}[h!]
	\begin{center}
		\begin{tikzpicture}[scale=1] 
			
			\shade[left color=green!10, right color=green!50] (0,0) rectangle (4,2) node[midway] {Newtonian theory};
			\shade[left color=green!50, right color=green!100] (4,0) rectangle (8,2) node[midway] {Special relativity};
			\shade[left color=red!20, right color=red!70] (8,0) rectangle (12,2) node[midway] {Tachyon theory};
			\shade[left color=red!70, right color=red!100] (12,0) rectangle (16.5,2) node[midway] {Post-Carrollian theory};

			\draw[->] (0,0) -- (17,0) node[right] {$v$};
			
			\draw[decorate, decoration={zigzag, segment length=6, amplitude=2}] (4,0) -- (4,2);
			\draw[decorate, decoration={zigzag, segment length=6, amplitude=2}] (11.9,0) -- (11.9,2);

			\node at (0,0) [below] {0};
			\node at (2,.75) [below] {$v\ll c$};
			\node at (6,.75) [below] {$v < c$};
			\node at (8,0) [below] {$c$};
			\node at (10,.75) [below] {$v > c$};
			\node at (14,.75) [below] {$v\gg c$};
			\node at (16.5,0) [below] {$\infty$};
		\end{tikzpicture}
		\caption{\small Theories of motion based on their velocities in comparison to the speed of light (see Table \ref{tab:example1}).\newline At $v=0$, the relations \eqref{teee}, \eqref{tmm} indicate that the particle possesses energy but zero momentum, characterizing a particle at rest. Conversely, at $v=\infty$, the relations \eqref{tee}, \eqref{tm} reveal that the particle carries momentum while having zero energy, corresponding to a magnetic Carroll particle.}\label{threthe}
	\end{center}
\end{figure}
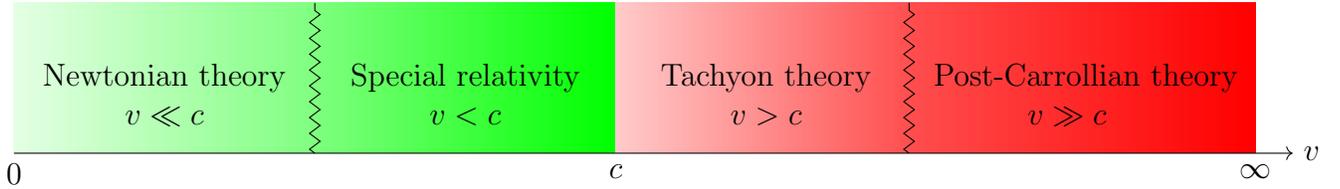

Therefore, in this work, we demonstrate that applying the ultra-relativistic limit\footnote{By the \textit{ultra-relativistic limit} in tachyon theory, we mean the regime where $\frac{v}{c} \gg 1$. This should not be confused with the \textit{ultra-relativistic limit} in special relativity, where $\frac{v}{c} \to 1$.} ($\frac{v}{c}\gg 1$) to tachyon theory leads to the formulation of a framework referred to as ``post-Carrollian theory''. Similarly, we recall that Newtonian theory emerges from the non-relativistic limit ($\frac{v}{c}\ll 1$) applied to special relativity. To compare the above expansions with those in special relativity refer to \eqref{teee}, \eqref{tmm}. Figure \ref{threthe} illustrates the area we are developing and its relationship to other established theories of motion. Green theories are linked to observable phenomena, while red theories remain yet hypothetical (see also Table \ref{tab:example1}).

Accordingly, we derive formulas for energy, momentum, energy-momentum relation, Newton's second law of motion, gravitational potential, and gravitational field within the post-Carrollian framework. When we derive the post-Carrollian energy-momentum relation, two applications arise. First, we use it to derive the (generalized) Carroll–Schr\"odinger equation, which has also been obtained through an alternative approach in \cite{Najafizadeh:2024imn}. Second, we explore the properties of a classical ideal gas composed of post-Carroll particles—objects that obey post-Carroll's laws of motion. Our focus is on the canonical ensemble, where we compute the thermodynamic quantities associated with this system.

As is well-known, the weak and static Einstein's field equations, when coupled to relativistic dust, can be reduced to the Poisson's equation for Newtonian gravity. Similarly, we derive the Poisson's equation for post-Carrollian gravity by studying the weak and static Einstein's field equations coupled to a tachyon dust. It finds that the gravitational force between post-Carroll particles is attractive, much like the force between ordinary particles. Finally, by applying the geodesic equation, we determine that the post-Carrollian gravitational field is radially outward, in contrast to the radially inward direction found in Newtonian gravity. This leads us to attribute a ``positive mass charge'' to Newtonian particles and assign a ``negative mass charge'' to post-Carrollian particles.

The paper is organized as follows: In Section \ref{CP}, we formulate post-Carrollian mechanics. In Section \ref{cig}, we investigate the properties of a post-Carrollian ideal gas. In Section \ref{CGFo1}, we explore post-Carrollian gravity by deriving both the post-Carrollian gravitational potential and the post-Carrollian gravitational field. Finally, Section \ref{conclu} presents the conclusions and discusses future directions. The appendices include additional information for the reader's convenience and to facilitate comparison with the main text. Appendix \ref{tom} provides a brief review of the known theories illustrated in Figure \ref{threthe}: (i) special relativity, (ii) Newtonian theory, and (iii) tachyon theory. Appendix \ref{hd} reproduces the generalized Carroll–Schr\"odinger equation. Finally, Appendix \ref{GGP1} details the derivation of both the Newtonian gravitational potential and the corresponding gravitational field.

\paragraph{Conventions:} We use the mostly plus signature for the Minkowski metric $\eta _{\mu \nu }={\text{diag}}(-1,+1,\ldots,+1)$. Greek indices $\mu$, $\nu$, $\ldots$ run over spacetime dimensions: $0,\ldots,d-1$, while Latin indices $i$, $j$, $k$ run over spatial dimensions: $1,\ldots,d$. Throughout the paper, we utilize shorthand notations: $\p_t:=\p/\p t$, $\p_x:=\p/\p x$, and $\p_i:=\p/\p x^i$. In addition, we define the d'Alembertian operator as $\Box:=-\,\frac{1}{c^2}\,\p_t^2+\p^i\p_i$. Throughout this work, the velocity vector $\mathbf{v}$ (with magnitude $v=|\mathbf{v}|$) represents the velocity of the test particle as measured in the observer's inertial frame. We emphasize that this is distinct from the relative velocity $\mathbf{u}$ between two inertial frames—a concept not discussed in our analysis.

\section{Post-Carrollian mechanics} \label{CP}

To make this section easier to follow, we have provided a detailed review of tachyons in Appendix \ref{Tachyons}. Utilizing this appendix, we derive expressions for energy, momentum, energy-momentum relation, and Newton's second law of motion within the post-Carrollian framework.

\subsection{Energy}

By expanding the tachyon Lorentz factor \eqref{gamma} in powers of $c$, one obtains $\gt=\pm\,\frac{c}{v}+\mathcal{O}(c^3)$. As a result, the tachyon energy \eqref{te} can be expressed as \eqref{tee}, where we consider only positive values. As discussed earlier, by assigning the vanishing leading term to the zero energy of magnetic Carroll particles, the subleading term yields 
\begin{align}
	\boxedB{\mathrm{E}\pc=\frac{m\,c^{\,3}}{v}\,.}\label{cen}
\end{align}
We refer to this as the ``post-Carrollian energy''. We observe that while the Newtonian energy \eqref{Gem}, $\mathrm{E}\nn=m v^2/2$, is proportional to $v^2$, the post-Carrollian energy \eqref{cen} is inversely proportional to the velocity magnitude $v$.

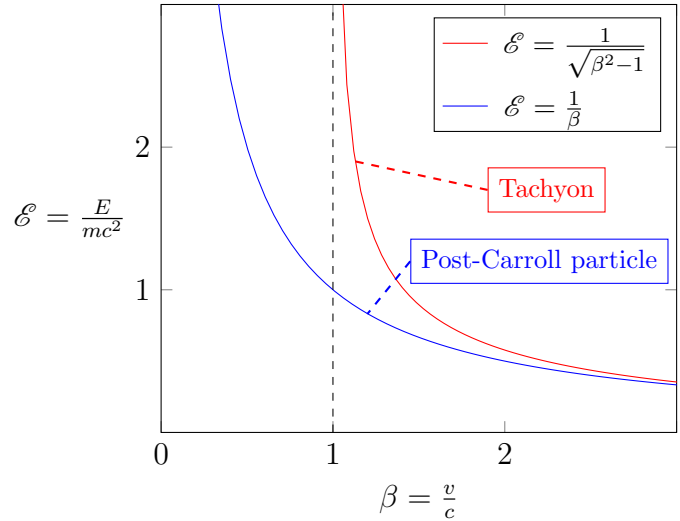
\begin{figure}[h!]
	\begin{center}
		\begin{tikzpicture}[scale=1]
			\begin{axis}[
				xlabel={$\beta=\frac{v}{c}$},
				ylabel={$\mathscr{E}=\frac{E}{mc^2}$},
				ylabel style={at={(ticklabel* cs:0.5)}, anchor=east, rotate=-90, right=-60pt},
				xmin=0,
				xmax=3,
				ymin=0,
				ymax=3,
				legend pos=north east,
				legend style={row sep=0.3cm, font=\small},
				xtick={0,1,2,3,4},
				xticklabels={0,1,2},
				ytick={1,2,3},
				yticklabels={1,2},
				]
				
				\addplot[color=red,domain=1:5,samples=100, restrict y to domain=0:5] {1/sqrt(x^2-1)};
				\addlegendentry{$~\mathscr{E} = \frac{1}{\sqrt{\beta^2-1}}$}
				
				\addplot[color=blue,domain=0:5,samples=100, restrict y to domain=0:5] {1/x};
				\addlegendentry{$~\!\!\!\!\!\!\!\!\!\!\!\!\!\mathscr{E} = \frac{1}{\beta}$}
				
				\addplot[color=black,dashed,domain=0:5,samples=2, restrict y to domain=0:5] coordinates {(1, 0) (1, 5)};
				
				\node[fill=white, draw=blue, anchor=west] at (axis cs: 1.45, 1.2) {\color{blue}\footnotesize{Post-Carroll particle}};
				\node[fill=white, draw=red, anchor=west] at (axis cs: 1.9, 1.7) {\color{red}{\footnotesize Tachyon}};
				
				\draw[dashed, thick, blue] (axis cs: 1.45, 1.2) -- (axis cs: 1.2, 0.83);
				\draw[dashed, thick, red] (axis cs: 1.9, 1.7) -- (axis cs: 1.13, 1.9);
				
			\end{axis}
		\end{tikzpicture}
	\end{center}
	\caption{\footnotesize This figure illustrates that the velocity of a tachyon is bounded within the range $c < v < \infty$, whereas the velocity of a post-Carroll particle is unbounded, ranging from $0 < v \leqslant \infty$. At high velocities ($v \gg c$), the post-Carroll particle curve coincides with the tachyon curve.}
	\label{TC}
\end{figure}

An important conclusion is found here. Whereas the velocity of a tachyon is confined within the range $c<v<\infty$, the velocity of a post-Carroll particle is unbounded, ranging from $0 < v \leqslant \infty$ (recall that $v=0$ corresponds to electric Carroll particles). Hence, the post-Carrollian energy \eqref{cen} can take on arbitrary values $0 \leqslant \mathrm{E}\pc<\infty$, allowing post-Carroll particles to have arbitrary velocity with arbitrary energy. In particular, when the velocity is infinite ($\frac{c}{v}\to 0$), the post-Carroll particle reduces to a magnetic Carroll particle with zero energy. The energy curve in terms of velocity, for both tachyons and post-Carroll particles, is displayed in Figure \ref{TC}. This can be compared with Figure \ref{f1}, illustrating the curve for relativistic particles and Newtonian particles (objects that obey Newton's laws of motion). 

\subsection{Momentum}

To derive the momentum formula in the post-Carrollian framework, we express the tachyon's total momentum \eqref{tp} as a series expansion in powers of $c/v$ (considering only positive values, since $\gt=\pm\,\frac{c}{v}+\mathcal{O}(c^3)$). This yields the total momentum of tachyon \eqref{tp} as (see equivalently \eqref{tm})
\be 
\mathbf{P}=mc\le(1\,+\,\frac{c^2}{2v^2}\,+\,\mathcal{O}\le(\,(c/v)^4\,\ri)\ri)\,\mathbf{\hat{v}}\,,\label{pp}
\ee
where $v = |\mathbf{v}| = \sqrt{v_i v_i}$ denotes the magnitude of the velocity vector $\mathbf{v}$, and $\mathbf{\hat{v}}=\mathbf{v}/v$ is the velocity unit vector. As discussed earlier, the leading term in the expansion \eqref{tm} or \eqref{pp} yields the magnetic Carroll particle momentum and the subleading term corresponds to what we identify as the ``post-Carrollian momentum''
\begin{align}
	\boxedB{\mathbf{P}\!\pc=\frac{1}{2}\,\frac{m\,c^{\,3}}{v^{\,2}}\,\mathbf{\hat{v}}\,.}\label{cmo}
\end{align}
As we observe, the post-Carrollian momentum \eqref{cmo}, which aligns with the direction of velocity, demonstrates an inverse relationship with the square of the velocity magnitude. This is in contrast to the Newtonian momentum \eqref{Gem}, $\mb{P}\!\nn=m\,\mb{v}$, which is directly proportional to the velocity magnitude.

An alternative perspective on the post-Carrollian momentum is as follows. By analogy with the Newtonian energy defined in \eqref{sube} as $\mathrm{E}\nn=\mathrm{E}-\mathrm{E}_0$, we introduce the post-Carrollian momentum through the relation
\be 
\mathbf{P}\!\pc=\mathbf{P}-\mathbf{P}_0\,.\label{cm}
\ee 
This is the total momentum of a tachyon $\mathbf{P}$ subtracted by the so-called ``rest momentum'' of the tachyon $\mathbf{P}_0=mc\,\mathbf{\hat{v}}$. Here, we use the term ``rest momentum'' in analogy with ``rest energy'' in the Newtonian case. The concept of rest momentum is further explained in Appendix \ref{Tachyons}. As we observe, the rest momentum $\mathbf{P}_0$ is nothing but the momentum of a magnetic Carroll particle, which has zero energy. Therefore, by omitting terms of order $\mathcal{O}\le((c/v)^4\ri)$ in \eqref{pp} and substituting the remaining terms into \eqref{cm}, we obtain the post-Carrollian momentum \eqref{cmo}, offering an alternative approach analogous to the Newtonian case. We note that when $v \to \infty$ ($\frac{c}{v}\to 0$), the post-Carrollian momentum \eqref{cmo} vanishes, $\mathbf{P}\!\pc=0$, and the total momentum in \eqref{cm} reduces to the rest momentum—the momentum of a magnetic Carroll particle.

\subsection{Energy-momentum relation}

Combining the above expressions for energy \eqref{cen} and momentum \eqref{cmo} and eliminating the velocity, we acquire the ``post-Carrollian energy-momentum relation''
\begin{align}
	\boxedB{|\mathbf{P}\!\pc|=\frac{\le(\mathrm{E}\pc\ri)^2}{2\,m\,c^{\,3}}\,.}\label{cem}
\end{align}
Here, $|\mathbf{P}\!\pc|=\sqrt{\mathrm{P}^i\!\!\pc\,\mathrm{P}_{_{\!\!\mathrm{PC}}i}}$ ($i=1,2,\ldots,d$) denotes the magnitude of the post-Carrollian momentum vector. Alternatively, this outcome could also be acquired directly from the tachyon energy-momentum relation \eqref{tem}, by considering the condition $\mathrm{E}\ll mc^2$ (which corresponds to $\mathbf{P}\!\pc\ll \mathbf{P}_0$), and applying \eqref{cm}, i.e. $|\mathbf{P}\!\pc|=|\mathbf{P}|-mc$.

In the following subsections, we present an application of the post-Carrollian energy-momentum relation \eqref{cem}. An additional application will be explored in Section \ref{cig}.

\subsubsection{Carroll–Schr\"odinger equation} 

The Carroll–Schr\"odinger equation, which describes an equation in two spacetime dimensions, was initially discovered using a technique detailed in \cite{Najafizadeh:2024imn}. Here, our goal is to reproduce this equation by applying the quantum mechanical prescription to the post-Carrollian energy-momentum relation. This derivation is similar to the traditional method of obtaining the Schr\"odinger equation \eqref{sch} from the Newtonian energy-momentum relation \eqref{Gemr}. To this end, let us consider the relation \eqref{cem} in one dimension
\be 
\mathrm{P}\!\pc=\frac{\le(\mathrm{E}\pc\ri)^2}{2\,m\,c^3}\,.\label{1d}
\ee
In order to formulate a quantum wave equation in the post-Carrollian framework, we apply the following quantum mechanical prescription
\be 
\mathrm{P}\!\pc\to i\hbar \,\p_x~, \quad\qquad \mathrm{E}\pc\to-\,i\hbar \,\p_t\,,\label{pre}
\ee
to the equation \eqref{1d}. This leads us to
\begin{align}
	{\le(i\hbar\,c\,\p_x+\frac{\hbar^2}{2m c^2}\,\p_t^2 \ri)\psi=0\,,}\label{cse}
\end{align} 	
which, by interpreting $\psi$ as a field, corresponds precisely to the obtained equation in \cite{Najafizadeh:2024imn}, called the Carroll–Schr\"odinger equation. This naming is motivated by the observation that under the substitution $x\leftrightarrow ct$, the equation \eqref{cse} transforms into the two-dimensional Schr\"odinger equation \eqref{sch}. In higher spacetime dimensions, the equation corresponding to \eqref{cse} was derived in \cite{Najafizadeh:2024imn} and called the ``generalized Carroll–Schr\"odinger equation''. The derivation of this generalized equation through the quantum mechanical prescription requires additional details, which are provided in Appendix \ref{hd}. Below, we examine the symmetry algebra associated with the Carroll–Schr\"odinger equation \eqref{cse}. We note that the symmetry algebra corresponding to the generalized Carroll–Schr\"odinger equation \eqref{inany} has yet to be determined.

\subsubsection{Carroll–Schr\"odinger algebra} 

In $1+1$ spacetime dimensions, the Carroll algebra \cite{Levy1965}, involving time translation $H$ (Hamiltonian), space translation $P$, and Carroll boost $B$, is given by $[\,P\,,\,B\,]=H$. Here, the Hamiltonian is a central charge and commutes with all generators. In particular, the relation $[\,H\,,\,B\,]=0$ has a physical interpretation: Carroll boosts leave the energy of any state invariant. This algebra can admit a central charge $M$, leading to the centrally extended Carroll algebra 
\begin{align}
	[\,P\,,\,B\,]=H\,,   \quad\qquad   [\,H\,,\,B\,]=M\,, \label{carrext}
\end{align}
which is satisfied by the representations
\be 
H=\p_t\,, \qquad B=x\,\p_t-i m\,t\,, \qquad P=\p_x\,, \qquad {M}=-\,i m\,.\label{repr}
\ee 
This structure was called the ``Carroll-Bargmann algebra'' in \cite{Najafizadeh:2024imn}, drawing an analogy to the Bargmann algebra, which is a central extension of the Galilei algebra. More recently, the reference \cite{Ecker:2025ncp} has been referred to \eqref{carrext} as the ``post-Carrollian algebra'', where the Hamiltonian $H$ is no longer a central charge. Physically, this implies that a boost applied to a post-Carrollian state alters its energy. This can be understood as incorporating corrections to the magnetic Carroll theory \cite{Ecker:2025ncp}.

Accordingly, while the leading terms in \eqref{tee}, \eqref{tm} (magnetic Carroll theory) correspond to the Carroll algebra, the subleading terms (post-Carrollian theory) are linked to the post-Carrollian algebra \eqref{carrext}. To see this connection, one should demonstrate that the equation \eqref{cse}, derived from the post-Carrollian energy-momentum relation \eqref{1d}, remains invariant under the transformation  
\be  
\delta\psi= \big(~\lambda_{_H}\,H~+~\lambda_{_P}\,P~+~\lambda_{_B}\,B~+~\lambda_{_M}\,M~\big)\,\psi  
\ee  
where $H$, $P$, $B$, and $M$ are the generators of the post-Carrollian algebra \eqref{carrext}, and $\lambda_{_H}$, $\lambda_{_P}$, $\lambda_{_B}$, $\lambda_{_M}$ denote the associated transformation parameters. This demonstration can be easily verified by rewriting the equation \eqref{cse} (assuming $c=1=\hbar$) in terms of generators, which yields $(iP+\frac{1}{2m}\,H^2)\,\psi=0$, where the representations in \eqref{repr} are utilized. It is then straightforward to see that this equation remains invariant under the transformations of $H$ and $P$. The invariance under the transformations $B$ and $M$, i.e. $\delta\psi=(\lambda_{_B}\,B+\lambda_{_M}\,M)\,\psi$, leads to the condition: $H\lambda_{_B}\big(i+\frac{M}{m}\big)\,\psi=0$. This condition is satisfied by setting $M=-\,im$, as specified in \eqref{repr}. As a result, while the Carroll algebra does not describe the symmetry of the equation \eqref{cse}, we have demonstrated that the post-Carrollian algebra \eqref{carrext} does indeed constitute its symmetry algebra.

It is worth noting that the post-Carrollian algebra (Carroll-Bargmann algebra) \eqref{carrext} can be conformally extended by including the generators of dilatation $D$ and the spatial special conformal transformation $\mathbb{K}$. The resulting algebra, characterized by the following nonzero commutation relations
\be
\begin{aligned}
	\begin{aligned}
		&[\,P\,,\,B\,]=H\,,       \quad & &[\,H\,,\,B\,]={M}\,,\\[2pt]
		&[\,P\,,\,D\,]=2\,P\,, \quad & &[\,D\,,\,\mathbb{K}\,]=2\,\mathbb{K}\,, \\[2pt]
		&[\,H\,,\,D\,]=H\,, \quad & &[\,P\,,\,\mathbb{K}\,]=D\,, \\[2pt]
		&[\,D\,,\,B\,]=B\,, \quad & &[\,H\,,\,\mathbb{K}\,]=B\,,\label{csch}
	\end{aligned}
\end{aligned}
\ee
is referred to as the ``Carroll–Schr\"odinger algebra'' \cite{Najafizadeh:2024imn} (see also \cite{Afshar:2024llh}), which has a dynamical critical exponent of $z=1/2$. This becomes isomorphic to the Schr\"odinger algebra (a conformal extension of the Bargmann algebra with $z=2$) upon interchanging the generators $P$ and $H$, aligning with the motivation behind its naming. It has been demonstrated \cite{Najafizadeh:2024imn} that the Carroll–Schr\"odinger algebra \eqref{csch} is the symmetry algebra of the Carroll–Schr\"odinger equation \eqref{cse}, analogous to how the Schr\"odinger algebra governs the symmetry of the Schr\"odinger equation.

Up to this point, we have focused on symmetries in $1+1$ dimensions. In higher dimensions, where rotation generators emerge, the post-Carrollian algebra \eqref{carrext} can be extended \cite{Ecker:2025ncp}. However, $M_i$ can no longer be central, as it carries a spatial vector index and thus does not commute with rotations. As a result, this extended algebra fails to serve as the symmetry algebra for equation \eqref{inany}, where the mass parameter $m$ lacks a vector index. Consequently, the extension of the Carroll–Schr\"odinger algebra \eqref{csch} to higher spacetime dimensions, or equivalently, the symmetries of the generalized Carroll–Schr\"odinger equation \eqref{inany}, remain unknown (see \cite{Najafizadeh:2024imn, Afshar:2024llh}), a problem we are currently investigating.

We note that extending the Carroll algebra by including a central charge—specifically, the generator $M$ in our case \eqref{carrext}—gives rise to moving massive Carroll particles, which we refer to as post-Carroll particles. This suggests that central extensions play a crucial role in enabling mobility within Carrollian frameworks. A further illustration appears in $1+2$ spacetime dimensions: the doubly extended Carroll algebra involves two central parameters, $A_i$ (with $i = 1, 2$), entering the commutators of momenta $[\,P_i,\, P_j\,] = \epsilon_{ij} A_1$ and boosts $[\,B_i,\, B_j\,] = \epsilon_{ij} A_2$. As shown in \cite{Marsot:2021tvq}, these extensions also result in moving Carroll particles, called exotic Carroll particles (see also \cite{Zeng:2024bcl} for an application to the Hall effect).

\subsection{Newton's second law}

Newton's second law of motion, which is a fundamental postulate and a law of nature, states that the time derivative of the momentum is equal to the force
\be 
\mb{F}=\frac{\mathrm{d}\mb{P}}{\mathrm{dt}}\,.\label{nsl}
\ee 
Utilizing this definition, and employing the post-Carrollian momentum vector $\mathbf{P}\!\pc$ derived in \eqref{cmo}, we find (assuming the direction of the velocity vector remains constant over time; i.e. $d\mathbf{\hat{v}}/dt=0$)
\be 
\boxedB{~~
	\begin{aligned}
		\mathbf{F}\!\pc=\frac{\mathrm{d}\mathbf{P}\!\pc}{\mathrm{dt}}\,
		=\frac{\mathrm{d}}{\mathrm{dt}}\le(\frac{1}{2}\,\frac{m\,c^{\,3}}{v^{\,2}}\,\mathbf{\hat{v}}\ri)&=-\,m\,\le(\frac{c}{v}\ri)^3\mathbf{a}\,,\label{tcforc} 
	\end{aligned}
	~~}
\ee
where $\mb{a}=\frac{d}{dt}\mathbf{v}=a\,\mathbf{\hat{v}}$ is the acceleration vector, with $a=dv/dt$ being the magnitude of the acceleration vector. We refer to \eqref{tcforc} as ``post-Carrollian Newton's second law of motion''.

This formula, which includes a negative sign, suggests the possibility of an outward gravitational field produced by post-Carrollian matter, as will be demonstrated later using the geodesic equation. The presence of velocity in the denominator of the momentum relation \eqref{cmo} is responsible for the emergence of the negative sign in the above expression. We note that this result depends on the magnitude of the particle's velocity $v$, distinguishing it from the Newtonian case, where $\mb{F}\nn = m\,\mb{a}$ \eqref{b4}.

Another approach to derive \eqref{tcforc} is as follows. By expanding the tachyon version of Newton's second law \eqref{tforc} in terms of powers of $c/v$, we get
\be 
\mb{F}=m\,\le(\frac{c}{v}\ri)\Big(~\mathbf{a}-(\mathbf{\hat{v}}\cdot{\mathbf{a}})\,\mathbf{\hat{v}}~\Big)+\frac{1}{2}\,m\,\le(\frac{c}{v}\ri)^3\Big(\,\mathbf{a}-\,3\,(\mathbf{\hat{v}}\cdot{\mathbf{a}})\,\mathbf{\hat{v}}\,\Big)\,+\,\mathcal{O}\le((c/v)^5\ri)\,.\label{force3}
\ee
Assuming that the acceleration vector and velocity vector are aligned $\mb{a}=a\,\mathbf{\hat{v}}$, we have $(\mathbf{\hat{v}}\cdot{\mathbf{a}})\,\mathbf{\hat{v}}=\mathbf{a}$. This eliminates the leading term $(c/v)^1$ in \eqref{force3} while simultaneously simplifying the subleading term $(c/v)^3$. Consequently, by dropping the higher order terms $\mathcal{O}\le((c/v)^5\ri)$, the relation \eqref{force3} reduces to $\mathbf{F} = -\,m\,\mathbf{a}\le({c}/{v}\ri)^3$. This provides another method for obtaining \eqref{tcforc}.

To close this section, let us consider the general case where the velocity direction varies with time $d\mathbf{\hat{v}}/dt\ne 0$. In this case, we find 
\be \mathbf{F}\!\pc=\frac{1}{2}\,m\,\le(\frac{c}{v}\ri)^3\Big(\,\mathbf{a}-\,3\,(\mathbf{\hat{v}}\cdot{\mathbf{a}})\,\mathbf{\hat{v}}\,\Big)=\frac{1}{2}\,m\,\le(\frac{c}{v}\ri)^3\Big(\,\mb{a}_{\!_\perp}-\,2\,\mb{a}_{\,\shortparallel}\Big)\,,\label{both}
\ee 
where $\mb{a}=\mb{a}_{\,\shortparallel} + \mb{a}_{\!_\perp}$ is separated into the part parallel to the velocity and the part perpendicular to it. As observed, the parallel part is the term derived earlier in \eqref{tcforc}.

\section{Post-Carrollian ideal gas} \label{cig}

This section presents a significant application of the post-Carrollian energy-momentum relation \eqref{cem}, examining the properties of an ideal gas composed of post-Carroll particles. 

\newpage

In the canonical ensemble, we consider a system consisting of $N$ identical post-Carroll particles, where there are no internal degrees of motion to be taken into account. The system is confined to a volume $V$ and is in thermal equilibrium at a temperature $T$. Since there are no internal interactions involved, the energy of the system is entirely kinetic. Therefore, utilizing the post-Carrollian energy-momentum relation \eqref{cem}, the energy of a single post-Carroll particle can be expressed as follows:
\be 
H_1(q,p)=\sqrt{2mc^3 |\mathbf{P}|}=\sqrt{2mc^3}~\sqrt[4]{p^ip_i}\,.\label{H1}
\ee
Hence, the one-particle partition function would be
\be 
Z_1=\frac{1}{h^3}\int d^3q~d^3 p ~e^{-\,\beta H_1}\,,\label{Z1}
\ee 
where $h$ is Planck constant and $\beta=1/(kT)$, with $k$ being Boltzmann's constant. By substituting \eqref{H1} into \eqref{Z1} and integrating over the entire phase space, we obtain the one-particle partition function as
\begin{align}
	Z_1~=~\frac{V}{h^3}\int_0^\infty (4\pi p^2 dp)~\exp\le(-\,\b c\,\sqrt{2mc\,p}\,\ri)~=~\frac{60 \,\pi\, V}{h^3\,m^4}\le(\frac{KT}{c^2}\ri)^6\,.
\end{align} 
Since the system is non-interacting, and hence the total energy is the sum of individual energies, one has
\be 
Z_N=\frac{1}{N!}\le(Z_1\ri)^N\,.
\ee 
As a result, using Stirling's formula $\ln(N!)=N\ln N-N$, we find the Helmholtz free energy
\be
A=-KT\ln Z_N=NKT\,\big(\ln\mathrm{X}-1\big)\,\label{hfe}
\ee
where
\be 
\mathrm{X}=\frac{N h^3 m^4 c^{12}}{60\,\pi V K^6 T^6}\,.
\ee
Using this, the complete thermodynamics of the post-Carrollian ideal gas can be 
derived in a straightforward manner. For instance, the chemical potential $\m$, entropy $S$, pressure $P$, internal energy $U$, the specific heat at constant volume $C_{_V}$, and the one at constant pressure $C_{_P}$, become
\be 
\boxedB{~~~
	\begin{aligned}
		&\m=KT\,\ln\mathrm{X}\,,            \quad & &U=6NKT\,,\\[6pt]
		&S=NK\,\big(\ln\mathrm{X}-7\big)\,, \quad & &C_{_V}=6NK\,, \\[6pt]
		&P=NKT/V\,,                         \quad & &C_{_P}=7NK\,.\label{cte}
	\end{aligned}
	~~~}
\ee
In addition, we find 
\be 
\boxed{~PV=\frac{1}{6}\,U\,,~}
\ee 
and the adiabatic index (the ratio of two specific heats) becomes
\be 
\boxed{~\ga=C_{_P}/C_{_V}=\frac{7}{6}\,.~} \label{adiaba}
\ee
We recall that for a classical (Newtonian) ideal gas, we have the relation $PV=\frac{2}{3}\,U$ and the adiabatic index $\gamma=\frac{5}{3}$. Besides, we have $PV=\frac{1}{3}\,U$ and $\gamma=\frac{4}{3}$ for an extreme relativistic gas \cite{Pathria:1996hda}. 


To close this section, let us establish a connection between our result for the adiabatic index of a post-Carrollian ideal gas \eqref{adiaba}, and the one for an ideal gas of Lifshitz particles, presented in Reference \cite{deBoer:2017ing}. Specifically, they found the adiabatic index to be 
\be 
\ga=C_{_P}/C_{_V}=1+\frac{z}{d}\,,\label{boer}
\ee
where $d$ represents the spatial dimensions and $z$ is the dynamical critical exponent, with $z\geq 1$. For $d=3$, this result aligns with that of a Newtonian ideal gas (with $z=2$) and an extreme relativistic gas (with $z=1$), as their values were previously mentioned. Importantly, we find that for $d=3$, the relation \eqref{boer} also agrees with our result in \eqref{adiaba} when $z=1/2$. This is the critical exponent linked to the Carroll–Schr\"odinger algebra, as explored in \cite{Najafizadeh:2024imn}. Consequently, this indicates that the findings in \cite{deBoer:2017ing} may also hold beyond the restriction $z\geq 1$, particularly for $z=1/2$, as demonstrated here. For comparison, the results of different gas types are listed in Table \ref{tab:sample_table22}.
\begin{table}[h!]
	\begin{center}
		\renewcommand{\arraystretch}{2.4} 
		\resizebox{1\textwidth}{!}{
			\begin{tabular}{|>{\centering\arraybackslash}m{5cm} 
					|>{\centering\arraybackslash}m{4cm} 
					|>{\centering\arraybackslash}m{4cm} 
					|>{\centering\arraybackslash}m{4cm}|}
				\hline
				\rowcolor{yell!15}\cellcolor{yell!15} Gas type  & Equation of state & Adiabatic index & Critical exponent \\
				\hline
				Classical ideal gas &  $\frac{PV}{U}=\frac{2}{3}$ & $\gamma=\frac{5}{3}$ & $z=2$  \\
				\hline
				Extreme relativistic gas  & $\frac{PV}{U}=\frac{1}{3}$ & $\gamma=\frac{4}{3}$ & $z=1$ \\
				\hline
				Post-Carrollian ideal gas  & $\frac{PV}{U}=\frac{1}{6}$ & $\gamma=\frac{7}{6}$ & $z=\frac{1}{2}$ \\
				\hline
		\end{tabular}}
		\caption{\footnotesize Comparison of different types of ideal gas.}
		\label{tab:sample_table22}
	\end{center}
\end{table}

\section{Post-Carrollian gravity} \label{CGFo1}

In this section, we aim to derive a post-Carrollian gravitational potential from Einstein's field equations in Section \ref{CGFo} and a post-Carrollian gravitational field from the geodesic equation in Section \ref{CGF}. To achieve this, we take the ultra-relativistic limit, analogous to the non-relativistic limit leading to the Newtonian framework. For the sake of comparison, the derivation of the Newtonian gravitational potential is included in Appendix \ref{GGP}, and the Newtonian gravitational field is detailed in Appendix \ref{ggfg}.

\subsection{Gravitational potential} \label{CGFo}

It is shown that the Einstein's field equations admit solutions for the energy-momentum tensor of tachyon dust (see e.g. \cite{Foster:1972wj,Srivastava:1977ay,Singh1985,Schwartz:1981qm,Schwartz:2019emp}). Accordingly, we utilize the Einstein's equations through the following steps: (i) applying a weak field approximation, (ii) assuming a time-independent gravitational metric, and (iii) imposing the ultra-relativistic limit on the tachyon dust. This process allows us to derive Poisson's equation for post-Carroll gravity and subsequently determine the corresponding gravitational potential. 

\newpage

According to the general relativity, the curvature of spacetime is related to the distribution of matter through the Einstein's equations
\be 
{R}^{\m\n}-\frac{1}{2}\,{R}\,g^{\m\n}=\frac{8\pi G\N}{c^4}~{T}^{\m\n}\,, \label{EE-}
\ee
where ${R}^{\m\n}$ is the Ricci tensor, $R$ the scalar curvature, $g^{\m\n}$ the inverse of the metric tensor, $c$ the speed of light, $G\N$ the Newtonian constant of gravitation, and $T^{\m\n}$ the energy-momentum tensor. Contracting both sides of \eqref{EE-} with the metric $g_{\m\n}$ in four  
dimensions, yields ${R}=-\frac{8\pi G\N}{c^4}~{T}$. Substituting this into \eqref{EE-}, we can rewrite the Einstein's equations \eqref{EE-} in an equivalent form
\be 
{R}^{\m\n}=\frac{8\pi G\N}{c^4} \,\Big({T}^{\m\n}-\frac{1}{2}\,\,{T}\,g^{\m\n} \Big)\,. \label{EE2-}
\ee 

\paragraph{(i) Weak field approximation:} 

Let us assume that the gravitational field is weak. This allows us to decompose the metric tensor into the Minkowski form plus a small perturbation: $ 
g_{\m\n}\approx\eta_{\m\n}+h_{\m\n}$. By substituting this expansion into the Einstein's equations \eqref{EE2-} and neglecting all terms of order $h^2$ and higher, we can derive the linearized Einstein's equations\footnote{Recalling that the Fronsdal equation \cite{Fronsdal:1978rb} represents the linearized higher spin equations, the left-hand side of \eqref{EE3-} is the Fronsdal equation with the helicity $s=2$.} for $h_{\mu\nu}$
\be 
\frac{1}{2}\,\Big( \p_\m \p^\a h_{\a\n}+ \p_\n \p^\a  h_{\a\m}-\p_\m \p_\n h-\,\p^\a \p_\a h_{\m\n}\Big)=\frac{8\pi G\N}{c^4}\, \Big({T}_{\m\n}-\frac{1}{2}\,\,{T}\,\eta_{\m\n} \Big)\,, \label{EE3-}
\ee
where $h=\eta^{\s\n}h_{\s\n}$. To ensure consistency with the contracted Bianchi identity, we note that the energy-momentum tensor on the right-hand side has to be computed to zeroth order in $h$, meaning it represents the unperturbed energy-momentum tensor of Minkowski spacetime and satisfies the standard conservation law $\partial^\nu T_{\mu\nu} = 0$.

The first three terms on the left-hand side of \eqref{EE3-} can be written as: $\p_\m\p^\a(h_{\a\n}-\frac{1}{2}\,\eta_{\a\n} \,h)$. It is then convenient to consider the so-called ``harmonic gauge'', or ``de Donder gauge''
\be 
\p^\a \Big(h_{\a\n}-\frac{1}{2}\,\eta_{\a\n} \,h\Big)=0\,.
\ee 
In this gauge, the first three terms in \eqref{EE3-} vanish, and the linearized Einstein's equations \eqref{EE3-} reduce to
\be 
-\,\p^\a \p_\a h_{\m\n}=\frac{16\pi G\N}{c^4}\, \Big({T}_{\m\n}-\frac{1}{2}\,\,{T}\,\eta_{\m\n} \Big)\,.\label{LE}
\ee

\paragraph{(ii) Time-independent metric:} We look now for solutions of the linearized equations \eqref{LE} by assuming that the spacetime geometry is time-independent, i.e. $\p_0 h_{\m\n}=0$. As a result, the linearized equations \eqref{LE} simplify to the ``weak and static field equations''
\be 
\nabla^2 h_{\m\n}=-\,\frac{16\pi G\N}{c^4}\, \Big({T}_{\m\n}-\frac{1}{2}\,\,{T}\,\eta_{\m\n} \Big)\,,\label{LE2}
\ee 
where $\nabla^2:=\p^i\p_i$ is the usual Laplace operator of three-dimensional Euclidean space. From this point on, we can proceed by employing an appropriate energy-momentum tensor in \eqref{LE2} for the problem at hand, whether it be Newtonian or post-Carrollian. The Newtonian case is comprehensively reviewed in Appendix \ref{GGP}. We follow a similar approach for the post-Carrollian case, as detailed below.

\paragraph{(iii) Ultra-relativistic limit:} 

Let us consider the weak and static field equations \eqref{LE2} and apply the energy-momentum tensor of a tachyon dust, as given by \cite{Foster:1972wj,Srivastava:1977ay,Singh1985}\footnote{The energy-momentum tensor of a tachyon perfect fluid is given by $T^{\m\n}=\le(\r+\frac{p}{c^2}\ri)U^\m U^\n-p\,g^{\m\n}$ \cite{CordeirodosSantos:2023zja}, where $\r$ is the tachyon proper mass density and $p$ is the pressure. Notably, this expression differs from the energy-momentum tensor of a conventional relativistic perfect fluid, $T^{\m\n}=\le(\r+\frac{p}{c^2}\ri)U^\m U^\n+p\,g^{\m\n}$, by a sign change in front of the pressure in the second term.
} 
\be 
T^{\m\n}=\r\,U^\m U^\n\,.\label{dustem}
\ee
Here, $\r>0$ represents the tachyon proper mass density and $U^\m$ is the spacelike four-velocity vector of the tachyon dust, given by (see e.g. \cite{Schwartz:2019emp})
\be 
U^\m=(c\,\gt, v^i \gt)\,,\label{4ve}
\ee 
with
\be
\gt=\frac{1}{\sqrt{v^2/c^2-1}}\,,\label{gamtt}
\ee
where $v^2=v(t)^2=v^i(t)v_i(t)$ can, in principle, be time-dependent. In fact, the spacelike four-velocity vector $U^\mu$ at any point of world line $x^\m(\tau)$ is defined as $U^\m={dx^\m}/{d\tau}$, where $x^\m = (ct,x^i)$ is the four-position and $\tau$ is the proper time, defined through $dt/d\tau=\gt$. More precisely, the components of \eqref{4ve} are $U^0=\frac{dx^0}{d\tau}=c\,\frac{dt}{d\tau} = c\,\gt$ and $U^i=\frac{dx^i}{d\tau}=\frac{dx^i}{dt}\,\frac{dt}{d\tau}= v^i\gt$, where $v^i={dx^i}/{dt}$.

We note that, as observed, relativistic particles and tachyons share the same algebraic form of the dust energy-momentum tensor, as seen in \eqref{dustem} and \eqref{emtr}. The key difference lies in the applied four-velocity vector: relativistic particles, characterized by a timelike 4-velocity vector \eqref{fourv}, satisfy $g_{\m\n}\,U^\m U^\n=-\,c^2$, whereas tachyons, with a spacelike 4-velocity vector \eqref{4ve}, satisfy $g_{\m\n}\,U^\m U^\n=c^2$.

The components of the energy-momentum tensor \eqref{dustem}, and its trace $T=g_{\m\n}\,T^{\m\n}\approx (\eta_{\m\n}+h_{\m\n}) T^{\m\n}$ up to zeroth order in $h$, are given by:
\begin{align}
	&T_{00}=\r\,c^2\,\gt^2\,\approx\,\r\,c^2\le(\frac{c}{v}\ri)^2\,, \label{T11}\\[8pt]
	&T_{i0}=\r\,c\,v^i\,\gt^2\,\approx\,\r\,c\,v^i\le(\frac{c}{v}\ri)^2\,, \label{T22}\\[8pt]
	&T_{ij}=\r\,v^i v^j\,\gt^2\,\approx\,\r\,v^i v^j\le(\frac{c}{v}\ri)^2\,, \label{T33}\\[8pt]
	&T=\eta_{\m\n}\,T^{\m\n}=\eta_{00}\,T^{00}+\eta_{ij}\,T^{ij}=\r\,\gt^2\le(-\, c^2+v^2\ri)=\r\,c^2\,.\label{TT}
\end{align}
The approximate symbols ($\approx$) in \eqref{T11}–\eqref{T33} indicate that the ultra-relativistic limit ($v\gg c$) has been applied, preserving post-Carrollian contributions, given that $\gt^2=c^2/v^2\,+\,\mathcal{O}(c^4)$.

Using these, we can examine different components of the weak and static field equations. For instance, by employing \eqref{T11} and \eqref{TT}, the $00$-component of the weak and static field equations \eqref{LE2}, up to the order of $\r \,G\N/c^2$, simplifies to
\be 
\nabla^2 h_{00}=-\,\frac{8\pi \r \,G\N}{c^2}\,.\label{jjj}
\ee 

\noindent By substituting 
\be 
h_{00}=-\,\frac{2\phi}{c^2}\,,\label{h0000}
\ee
into the equation \eqref{jjj}, we find Poisson's equation for post-Carrollian gravity 
\be 
\boxed{~\nabla^2 \phi={4\pi \r\,G\N}\,,\label{Poisson-}~}
\ee  
which, interestingly, is similar to the Newtonian case \eqref{Poisson}. The solution to this equation yields the ``post-Carrollian gravitational potential'' 
\be 
\boxed{~~\phi=-\,\frac{M G\N}{r}\,,~~} \label{key}
\ee  
where $M>0$ is the mass of post-Carrollian body. Therefore, post-Carrollian bodies share the same gravitational potential as Newtonian bodies. Importantly, it finds that the universal gravitational constant $G\N$ in the post-Carrollian framework has the same value as in the Newtonian framework. This is because the gravitational potentials, \eqref{key} and \eqref{Poisson}, in both frameworks have been derived from the same fundamental equation—Einstein's field equations \eqref{EE-}—which inherently include the gravitational constant $G\N$.

As a result, the gravitational energy potential $U$ and the gravitational force $\mathbf{F}$ within the post-Carrollian framework become the same as in the Newtonian case \eqref{Gu}
\be 
\boxed{~~U=m\,\phi=-\,\frac{m\,M\,G\N}{r}\,,\qquad\qquad \mathbf{F}=-\,\nabla U=-\,\frac{m\,M\,G\N}{r^2}\,\mathbf{\hat{r}}\,.~~}\label{cgf}
\ee 
Therefore, our analysis demonstrates that the gravitational force between two post-Carrollian bodies of masses $m$ and $M$ is attractive.

This attractive feature can also be understood from another approach. Raychaudhuri \cite{Raychaudhuri:1974ew} suggested that there is an attractive gravitational force between tachyons themselves. Moreover, as demonstrated in this work, tachyon theory and post-Carrollian theory are connected through a limit. Hence, it is natural to assume\footnote{This assumption also applies to relativistic particles and Newtonian particles, as they are connected through a limiting process and share the same attractive gravitational force.} that tachyons and post-Carroll particles share the same gravitational properties, indicating an attractive gravitational force between post-Carroll particles as well. This reasoning aligns with our findings, which are further supported by the formula presented in \eqref{cgf}.

By utilizing \eqref{T22} and \eqref{TT}, the $i0$-component of the weak and static field equations \eqref{LE2} does not include terms of the order $\r\,G\N/c^2$. Consequently, we have $\nabla^2 h_{i0}=0$ which demonstrates that 
\be h_{i0}=0\,,\label{hi0}\ee 
or $g_{i0}=\eta_{i0}+h_{i0}=0$.

In addition, using \eqref{T33} and \eqref{TT}, the $ij$-component of the weak and static field equations \eqref{LE2}, up to the order of $\r\,G\N/c^2$, can be expressed as
\be 
\nabla^2 h_{ij}= -\,\frac{8\pi \r\,G\N}{c^2}\le(2\,\frac{v_i \,v_j}{v^2}\,-\,\delta_{ij}\ri)\,.
\ee 
Assuming that velocities are independent of position, this equation leads to 
\be 
h_{ij}=-\,\frac{2\phi}{c^2}\,\le(2\,\frac{v_i v_j}{v^2}\,-\,\delta_{ij}\ri)\,.\label{hij}
\ee
Consequently, using \eqref{h0000}, \eqref{hi0} and \eqref{hij}, we get
\be
\boxedB{~~~
	\begin{aligned}
		g_{00}&=-\,1+h_{00}\,,\\[8pt]
		g_{i0}&=0\,,\\[2pt]
		g_{ij}&=(1-h_{00})\,\delta_{ij}+\,2\,\frac{v_i\, v_j}{v^2}\,h_{00}\,.\label{gha}
	\end{aligned}
	~~~}
\ee
Here, we observe a difference in the form of the metric component $g_{ij}$ compared to the Newtonian case \eqref{gmn}, while the other components of the metric remain the same in both frameworks.

As a result, in the post-Carrollian framework, the full line-element that solves the linearized Einstein's equations and describes the geometry generated by a weak and static field is given by
\be
\boxedB{~~~
	\begin{aligned}
		ds^2=g_{\m\n}dx^\m dx^\n=-\le(1+\frac{2\phi}{c^2}\ri)\,c^2 dt^2+\le(\delta_{ij}+\,\frac{2\phi}{c^2}\le(\delta_{ij}\,-\,2\,\frac{v_i\, v_j}{v^2}\ri)\ri)\,dx^i dx^j\,,
	\end{aligned}
	~~~}
\ee
where $\phi$ is a solution of the Poisson's equation \eqref{Poisson-}, given by \eqref{key}.

\subsection{Gravitational field} \label{CGF}
The Newtonian gravitational field can be derived from the geodesic equation under some assumptions. For the reader's convenience and to maintain a consistent approach throughout this section, we have provided a review of this derivation in Appendix \ref{ggfg}. In this section, we similarly aim to derive the post-Carrollian gravitational field using the geodesic equation. As our analysis demonstrates, this field is oriented radially outward.

The trajectory of test particles within the gravitational field of a curved spacetime is determined by the geodesic equation which applies equally to both timelike particles (relativistic particles) and spacelike particles (tachyons). The geodesic equation is given by
\be 
\frac{d^2x^{\mu}}{d\tau^2} + \Gamma^{\mu}_{\nu\rho}\frac{dx^{\nu}}{d\tau}\frac{dx^{\rho}}{d\tau} = 0\,,\label{geo} 
\ee 
or (when separating the summing indices into $0$ and $i$)
\be 
\frac{d^2x^{\mu}}{d\tau^2} 
+\Gamma^{\mu}_{ij}~\frac{dx^{i}}{d\tau}\frac{dx^{j}}{d\tau}
+2\,c\,\Gamma^{\mu}_{0i}~\frac{dt}{d\tau}\frac{dx^{i}}{d\tau}
+c^2\,\Gamma^{\mu}_{00}~\frac{dt}{d\tau}\frac{dt}{d\tau} = 0\,,\label{geo2} 
\ee
where $x^\m=(ct,x^i)$, $\tau$ represents the proper time, and $\Gamma^{\mu}_{\nu\rho}$ are Christoffel symbols
\be 
\Gamma^{\mu}_{\nu\rho}=\frac{1}{2}\,g^{\m\lambda}\Big({\p_\nu} g_{\rho\lambda}+\p_\rho g_{\nu\lambda}-\p_\lambda g_{\nu\rho}\Big)\,.\label{crif}
\ee  

\newpage

\noindent To apply the geodesic equation to tachyons, we use $dt/d\tau=\gt$, where $\gt$ is defined in \eqref{gamtt}. By applying the chain rule, we can express
\begin{align}
	\frac{dx^{\mu}}{d\tau}&=\gt\,\frac{dx^{\mu}}{dt}\,,\label{chain21}\\[8pt]
	\frac{d^2x^{\mu}}{d\tau^2}&=\gt^2\,\frac{d^2x^{\mu}}{dt^2}-\frac{\gt^4}{c^2}\,(\mathbf{v}\cdot\mathbf{a})\,\frac{dx^{\mu}}{dt}\,,\label{chain2}
\end{align}
where 
\be 
\frac{d\gt}{dt}=\frac{d\gt}{dv^i}\,\frac{dv^i}{dt}=-\,\frac{\gt^3}{c^2}\,(\mathbf{v}\cdot\mathbf{a})\,,
\ee
has been used, with $\mathbf{v} \cdot \mathbf{a}$ denoting the inner product of the velocity vector $v^i=dx^i/dt$ and the acceleration vector $a^i=dv^i/dt$ of the test particle. Consequently, using \eqref{chain21} and \eqref{chain2}, the geodesic equation \eqref{geo2} can be rewritten as
\be 
\gt^2~\frac{d^2x^{\mu}}{dt^2}-\frac{\gt^4}{c^2}\,(\mathbf{v}\cdot\mathbf{a})\,\frac{dx^{\mu}}{dt} 
+\gt^2~\Gamma^{\mu}_{ij}~v^i\,v^j
+2\,c\,\gt^2~\Gamma^{\mu}_{0i}~v^i
+c^2\,\gt^2~\Gamma^{\mu}_{00}= 0\,.\label{geo22} 
\ee
We note that if the magnitude of the velocity $v$ is time-independent, $\gt$ would become a constant factor (see \eqref{gamtt}), allowing $\gt^2$ to be factored out and dropped from each term in \eqref{geo22}. However, we retain this factor in every term of the equation, as we generally assume that velocity magnitude is time-dependent.

Similar to the Newtonian case, here, we define the post-Carrollian limit of the geodesic equation \eqref{geo}/\eqref{geo22} by imposing three assumptions:  

\noindent ($i$) {Weak gravitational field:} The gravitational field is treated as a small perturbation to the flat spacetime metric, $g_{\m\n}\approx\eta_{\m\n}+h_{\m\n}$. Consequently, the components derived in \eqref{h0000}, \eqref{hi0}, and \eqref{hij} should be applied.  

\noindent ($ii$) {Static gravitational field:} The gravitational field does not vary with time, implying $\p_0 g_{\m\n}=0$.  

\noindent ($iii$) {Ultra-relativistic limit:} The test particle's velocity is much larger than the speed of light, $v\gg c$.

The first assumption simplifies the Christoffel symbols to $\Gamma^{\mu}_{\nu\rho}=\frac{1}{2}\,\eta^{\m\lambda}({\p_\nu} h_{\rho\lambda}+\p_\rho h_{\nu\lambda}-\p_\lambda h_{\nu\rho})$. Next, using the components $h_{i0}=0$ and $h_{ij}=h_{00}\,(\,2\,\frac{v_i v_j}{v^2}\,-\,\delta_{ij}\,)$, as derived in \eqref{hi0} and \eqref{hij}, and incorporating the second assumption, $\p_0 \,h_{\m\n}=0$, along with the assumption that velocities are independent of position, the components of the Christoffel symbols can be expressed as follows:
\begin{align}
	&\Gamma^{0}_{ij}=\Gamma^{i}_{0j}=\Gamma^{0}_{00}=0\,,\nonumber\\[5pt]
	&\Gamma^{0}_{0i}=\Gamma^{i}_{00}=-\,\frac{1}{2}\,\p_i\, h_{00}\,,\nonumber\\[5pt]
	&\Gamma^{k}_{ij}=\frac{1}{2}\,\le[\le(\frac{2\,v_jv_k}{v^2}-\d_{jk}\ri)\!\p_i+\le(\frac{2\,v_iv_k}{v^2}-\d_{ik}\ri)\!\p_j-\le(\frac{2\,v_iv_j}{v^2}-\d_{ij}\ri)\!\p_k\ri]h_{00}\,.\label{components}
\end{align}
Using these components, the geodesic equation \eqref{geo22} for $\m=0$ leads to the following condition
\be
\boxedB{~~~
	\begin{aligned}
		-\,\frac{\gt^4}{c^2}\,(\mathbf{v}\cdot\mathbf{a})=\gt^2\,v^i\,\p_i h_{00}\,. \label{cond2}
	\end{aligned}
	~~~}
\ee
On the other hand, using \eqref{components}, we obtain
\be 
\Gamma^{k}_{ij}~v^i\,v^j=v^k\,v^i\,\p_i\,h_{00}-\,\frac{1}{2}\,v^2\,\p_k\,h_{00}\,,
\ee 
and thus, for $\m=k$, the geodesic equation \eqref{geo22} yields
\be 
\gt^2~	\frac{d^2x^{k}}{dt^2}-\,\frac{\gt^4}{c^2}\,(\mathbf{v}\cdot\mathbf{a})\,v^k 
+\gt^2~v^k\,v^i\,\p_i h_{00}-\,\frac{\gt^2}{2}\,v^2\,\p_k h_{00}-\,\frac{\gt^2}{2}\,c^2\,\p_k h_{00}=0\,. \label{geo33}
\ee 
By applying the condition \eqref{cond2}, the third term in the above equation can be expressed in terms of the second term, allowing \eqref{geo33} to simplify to (considering $h_{00}=-\,\frac{2\phi}{c^2}$)
\be
\gt^2~\frac{d^2x^{k}}{dt^2}-2\,\frac{\gt^4}{c^2}\,(\mathbf{v}\cdot\mathbf{a})\,v^k+\gt^2\le(1+\frac{v^2}{c^2}\ri)\p_k \phi=0\,.\label{ge2}
\ee

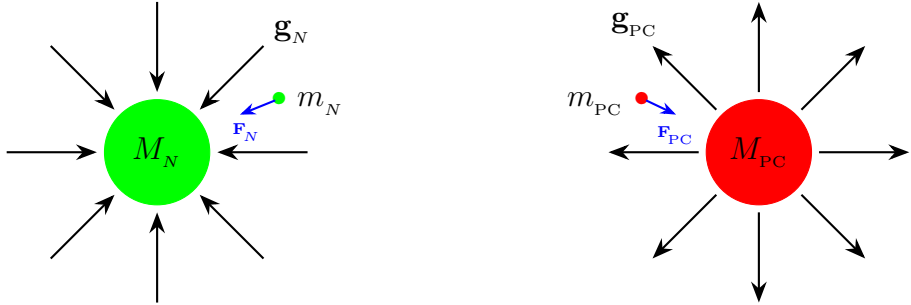
\begin{figure}[h!]
	\begin{center}
		\begin{tikzpicture}
			
			
			\filldraw[green] (-5,0) circle (0.7cm) node {\color{black}$M\nn$};
			\filldraw[red] (3,0) circle (0.7cm) node {\color{black}$M\pc$};
			
			\hspace{-5cm}\foreach \angle in {0, 45, 90, 135, 180, 225, 270, 315} {
				\draw[-{Stealth[scale=1.2]}, thick] (\angle:2) -- (\angle:0.8);
			}
			
			\node at (1.8, 1.6) {$\mathbf{g}\nn$};
			\node at (2, .65)   {${\color{green}\bullet}$~$m\nn$};
			
			\draw[-{Stealth[scale=.9]}, thick, blue] (1.57, .69) -- (1.1, .5) node[pos=.8,below] {{\tiny$\mathbf{F}_{_{\!{N}}}$}};
			
			\hspace{8cm}\foreach \angle in {0, 45, 90, 135, 180, 225, 270, 315} {
				\draw[-{Stealth[scale=1.2]}, thick] (\angle:0.8) -- (\angle:2);
			}
			
			
			\node at (-1.65, 1.7) {$\mathbf{g}\pc$};
			\node at (-2, .65) {$m\pc$~${\color{red}\bullet}$};
			
			\draw[-{Stealth[scale=.9]}, thick, blue] (-1.5, .69) -- (-1.1, .5) node[pos=1,below] {{\tiny$\mathbf{F}_{_{\!\mathrm{PC}}}$}};
			
		\end{tikzpicture}
		\caption{\footnotesize This figure illustrates that the Newtonian gravitational field $\mathbf{g}\nn$ generated by Newtonian matter $M\nn$ is radially inward, while the post-Carrollian gravitational field $\mathbf{g}\pc$ produced by post-Carrollian matter $M\pc$ is radially outward. A Newtonian particle of mass $m\nn$ (with $m\nn\ll M\nn$) in the gravitational field $\mathbf{g}\nn$ experiences an attractive force $\mathbf{F}\!\nn$, aligned with the gravitational field $\mathbf{g}\nn$. Similarly, a post-Carrollian particle of mass $m\pc$ (with $m\pc\ll M\pc$) in the gravitational field $\mathbf{g}\pc$ experiences an attractive force $\mathbf{F}\!\pc$, but anti-aligned with the gravitational field $\mathbf{g}\pc$.}
		\label{fig:grav_field}
	\end{center}
\end{figure}

At this stage, we apply the ultra-relativistic limit ($v\gg c$). In this regime, $\gt^2\approx c^2/v^2$ and $\gt^4\approx c^4/v^4$, implying that the first and second terms in \eqref{ge2} become of the same order and remain significant. Consequently, the term involving $\mathbf{v}\cdot\mathbf{a}$ does not vanish in the ultra-relativistic limit and plays a significant role. This behavior contrasts with the Newtonian case \eqref{ge}, where the $\mathbf{v}\cdot\mathbf{a}$ term ignores in the non-relativistic limit. As a result, in the ultra-relativistic limit only the first term in the parentheses becomes negligible, and thus, the equation \eqref{ge2} takes the form 
\be
\boxedB{~~~
	\begin{aligned}
		\frac{c^2}{v^2}\le[~\frac{d^2x^{k}}{dt^2}-2\,\le(\frac{v_k\,v_i}{v^2}\ri)\frac{d^2x^{i}}{dt^2}~\ri]
		=-\,\p_k \phi\,.\label{ge22}
	\end{aligned}
	~~~}
\ee	
This equation generally depends on the directions of velocity and acceleration, indicating a situation with velocity-dependent forces. However, for simple radial motion, we can consider the case where velocity and acceleration are aligned, corresponding to setting $i=k$ in \eqref{ge22}. In this case, the equation \eqref{ge22} reduces to
\be 
\frac{c^2}{v^2}~\frac{d^2x^{k}}{dt^2}=\p_k \phi\,.\label{ge222}
\ee  
If we now define the post-Carrollian gravitational field ${g}^k\pc$ as\footnote{One may define the post-Carrollian gravitational field as ${g}^k\pc:=\frac{d^2x^{k}}{dt^2}$, such that $\mathbf{g}\pc=\,\frac{v^2}{c^2}\nabla \phi$. However, this definition results in a velocity-dependent gravitational force, which is not consistent with our findings in \eqref{cgf}.} 
\be 
{g}^k\pc:=\frac{c^2}{v^2}\,\frac{d^2x^{k}}{dt^2}\,, \label{def}
\ee
the equation \eqref{ge222} leads to the following gravitational field equation in the post-Carrollian framework\footnote{In the case where velocity and acceleration of the test particle are perpendicular (as in circular motion), the term $\mathbf{v}\cdot\mathbf{a}$ in \eqref{ge2} vanishes. As a result, in the ultra-relativistic limit, the equation \eqref{ge2} simplifies to $\frac{c^2}{v^2}~\frac{d^2x^{k}}{dt^2}=-\,\p_k \phi$. Using the definition \eqref{def}, this further reduces to $\mathbf{g}\pc=-\,\nabla \phi$, resembling the Newtonian form. This is expected, as the perpendicular part in \eqref{both}, unlike the parallel part, appears with a positive sign.}
\be 
\boxedB{~~
	\begin{aligned}
		\mathbf{g}\pc=\nabla \phi\,.  \label{geod5}
	\end{aligned}
	~~}
\ee
\newpage
\noindent This result differs by a sign from the Newtonian case \eqref{GG}, where $\mathbf{g}\nn=-\,\nabla \phi$. Specifically, if we consider the post-Carrollian gravitational potential given in equation \eqref{key}, $\phi= -\,{M G\N}/{r}$, and substitute it into \eqref{geod5}, the ``post-Carrollian gravitational field'' becomes
\be 
\boxedB{~~
	\mathbf{g}\pc= \frac{M G\N}{r^{\,2}}~\mathbf{\hat{r}}\,. \label{geod51}
	~~}
\ee
Surprisingly, in contrast to the Newtonian case \eqref{GG}, this demonstrates that the post-Carrollian gravitational field $\mathbf{g}\pc$ generated by post-Carrollian matter of mass $M$ is directed radially outward. This is further illustrated by comparing it to the Newtonian case in Figure \ref{fig:grav_field}.

\paragraph{Mass charge:}

The attractive gravitational force between two Newtonian particles is given by \eqref{Gu}, while for two post-Carroll particles, it is provided in \eqref{cgf}. Since both expressions share the same mathematical form, we distinguish them by labeling the force and matter type: ``${N}$'' for the Newtonian framework and ``$\mathrm{PC}$'' for the post-Carrollian framework. Accordingly, the labeled versions of relations in \eqref{Gu}, \eqref{cgf} and \eqref{GG}, \eqref{geod51} are as below:
\begin{align} 
	\mathbf{F}\nn&=-\,\frac{G\N\,M\nn\,m\nn}{r^2}~\mathbf{\hat{r}}\,,\qquad\qquad
	\mathbf{F}\pc=-\,\frac{G\N\,M\pc\,m\pc}{r^2}~\mathbf{\hat{r}}\,,\label{gcf}\\[10pt]
	\mathbf{g}\nn&=-\,\frac{G\N\,M\nn}{r^2}~\mathbf{\hat{r}}\,,\qquad\qquad\qquad
	\mathbf{g}\pc=+\,\frac{G\N\,M\pc}{r^2}~\mathbf{\hat{r}}\,.\label{GGg}
\end{align}  
By relating these relations, we find: 
\be 
\boxedB{~~
	\begin{aligned}
		\mathbf{g}\nn=+~\frac{\mathbf{F}\nn}{m\nn}\,,\qquad\qquad\qquad
		\mathbf{g}\pc=-~\frac{\mathbf{F}\pc}{m\pc}\,.\label{charge}
	\end{aligned}
	~~}
\ee
Consequently, in comparison to electrostatics, it can be concluded that a ``positive mass charge'' can be attributed to the Newtonian particle $m\nn$, while a ``negative mass charge'' can be assigned to the post-Carrollian particle $m\pc$. Moreover, these relations demonstrate that $\mathbf{g}\nn$ and $\mathbf{F}\nn$ are aligned, whereas $\mathbf{g}\pc$ and $\mathbf{F}\pc$ are anti-aligned, as illustrated in Figure \ref{fig:grav_field}.

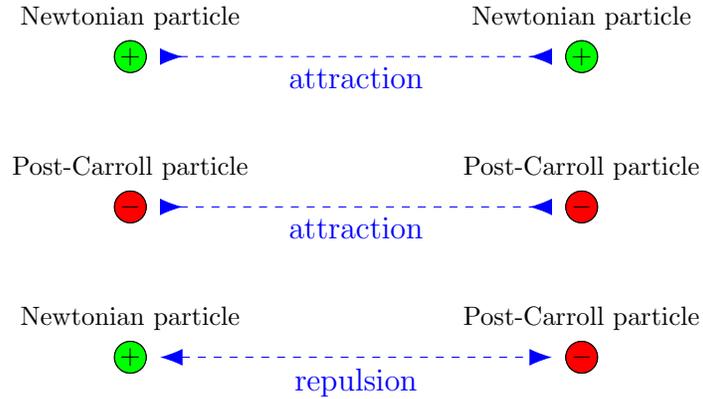
\begin{figure}[h!]
	\begin{center}
		\begin{tikzpicture}
			\tikzset{
				particle/.style={circle, draw=black, inner sep=0pt, minimum size=12pt},
				galilei/.style={particle, fill=green},
				carroll/.style={particle, fill=red},
				attraction/.style={{Latex[length=3mm,reversed]}-{Latex[length=3mm,reversed]},dashed,blue,shorten <=5pt, shorten >=5pt},
				repulsion/.style={{Latex[length=3mm]}-{Latex[length=3mm]},dashed,blue,shorten <=5pt, shorten >=5pt},
			}
			
			\node[galilei,label=above:{\footnotesize Newtonian particle}] (G1) at (0,0) {};
			\node[galilei,label=above:{\footnotesize Newtonian particle}] (G2) at (6,0) {};
			\node[carroll,label=above:{\footnotesize Post-Carroll particle}] (C1) at (0,-2) {};
			\node[carroll,label=above:{\footnotesize Post-Carroll particle}] (C2) at (6,-2) {};
			\node[galilei,label=above:{\footnotesize Newtonian particle}] (G3) at (0,-4) {};
			\node[carroll,label=above:{\footnotesize Post-Carroll particle}] (C3) at (6,-4) {};
			
			\draw[attraction] (G1) -- (G2) node[midway, below] {attraction};
			\draw[attraction] (C1) -- (C2) node[midway, below] {attraction};
			\draw[repulsion] (G3) -- (C3) node[midway, below] {repulsion} ;
			
			\node[galilei,label=above:] (G1) at (0,0) {\footnotesize $+$};
			\node[galilei,label=above:] (G2) at (6,0) {\footnotesize $+$};
			\node[carroll,label=above:] (C1) at (0,-2) {\footnotesize $-$};
			\node[carroll,label=above:] (C2) at (6,-2) {\footnotesize $-$};
			\node[galilei,label=above:] (G3) at (0,-4) {\footnotesize $+$};
			\node[carroll,label=above:] (C3) at (6,-4) {\footnotesize $-$};
			
		\end{tikzpicture}
		\caption{The gravitational force behavior of different types of particles.}\label{3333}
	\end{center} 
\end{figure}

Based on these observations, we can conclude that particles of different types—a Newtonian particle and a post-Carrollian particle—may experience a repulsive gravitational force, which can be expressed as: $\mathbf{F}=+\,{G\N~m\nn\,m\pc}~\mathbf{\hat{r}}/r^2$\,. This conclusion is further supported by Raychaudhuri's study \cite{Raychaudhuri:1974ew}, which reveals the existence of a repulsive gravitational force between tachyons and ordinary matter. Given that post-Carrollian theory is derived from tachyon theory, we can draw a parallel and infer that a similar repulsive interaction exists between post-Carrollian particles and ordinary matter (Newtonian particles). Therefore, the gravitational force behavior of different types of matter can be summarized in Figure \ref{3333}. Notably, the behavior of mass charge types in this domain—where like charges attract and opposite charges repel—is opposite to the behavior of charge types in electrostatics, where like charges repel and opposite charges attract.

\section{Conclusions and outlook}\label{conclu}

In this work, we aimed to develop a classical framework for post-Carrollian mechanics. To achieve this, we derived expressions for energy, momentum, energy-momentum relation, and Newton's second law of motion within the post-Carrollian framework. These expressions were obtained from the tachyon theory\footnote{The connection between tachyons and Carroll symmetry has also been studied in \cite{Gibbons:2002tv}.}, in a similar manner to how classical Newtonian mechanics can be obtained from special relativity. These findings are summarized in the last row of Table \ref{tab:example1}, along with the well-established formulas from other frameworks for comparison.
\begin{table}[t!]
	\begin{center}
		\setlength{\arrayrulewidth}{0.2mm}
		\resizebox{1\textwidth}{!}{
			
			\begin{tabular}{|p{3cm}|c|c|c|c|c|}
				\hline
				\rowcolor{yell!15}\cellcolor{yell!15}&{\footnotesize Energy}& {\footnotesize Momentum} &{\footnotesize Energy-momentum}&{\footnotesize Newton's second law}&{\footnotesize Quantum equation}\\[8pt] \hline
				\cellcolor{green!80}{\makecell{ \scriptsize{Newtonian theory}\,\\[5pt] $(v \ll c)$}}&
				\hyperref[Gem]{\color{black}$\mathrm{E}=\frac{1}{2}\,mv^2$} & \hyperref[Gem]{\color{black}$\mathbf{P}=m\mathbf{v}$} &  \hyperref[Gemr]{\color{black}$\mathrm{E}=\frac{\mathbf{P}^2}{2m}$} & \hyperref[b4]{\color{black}$\mathbf{F}=m\,\mathbf{a}$} & \hyperref[sch]{\color{black}$\le(i\hbar\p_t+\frac{\hbar^2}{2m}\nabla^2\ri)\!\psi=0$} \\[8pt] \hline
				\cellcolor{green!80}{\makecell{ {\scriptsize Special relativity}\\[5pt] $(v < c)$}}
				& \hyperref[p,e]{\color{black}$\mathrm{E}=\gamma\, m c^2$} & \hyperref[p,e]{\color{black}$\mathbf{P}=\gamma \,m \mathbf{v}$} & \hyperref[Rem]{\color{black}$\!\mathrm{E}^2=\mb{P}^2 c^2+m^2 c^4\!$} & \hyperref[rf]{\color{black}$\mathbf{F}=\ga^3\,m\,\mb{a}$} & \hyperref[Rqe]{\color{black}$\le(\Box-\frac{m^2 c^2}{\hbar^2}\ri)\!\phi=0$} \\ [8pt] \hline
				\cellcolor{red!80}{\makecell{ {\scriptsize Tachyon theory}\\[5pt] $(v > c)$}} & 
				\hyperref[te]{\color{black}$\mathrm{E}=\gt m c^2$} & \hyperref[tp]{\color{black}$\mathbf{P}=\gt m \mathbf{v}$} & \hyperref[tem]{\color{black}$\!\mathrm{E}^2=\mb{P}^2 c^2-m^2 c^4\!$} & \hyperref[tforc]{\color{black}$\mathbf{F}=-\,\gt^3\,m\,\mb{a}$} & \hyperref[tqe]{\color{black}$\le(\Box+\frac{m^2 c^2}{\hbar^2}\ri)\!\phi=0$} \\[8pt] \hline
				\cellcolor{red!80}{\makecell{ {\scriptsize Post-Carrollian theory}\\[5pt] $(v \gg c)$}}
				& \hyperref[cen]{\color{black}$\mathrm{E}=\frac{mc^3}{v}$} & \hyperref[cmo]{\color{black}$\mathbf{P}=\frac{mc^3}{2v^3}\,\mathbf{v}$} & \hyperref[cem]{\color{black}$|\mathbf{P}|=\frac{\mathrm{E}^2}{2mc^3}$} & \hyperref[tcforc]{\color{black}$\mathbf{F}=-\,m\le(\frac{c}{v}\ri)^3\mathbf{a}$} & \hyperref[inany]{\color{black}$\le(i\hbar c\nabla_{\!x}+\frac{\hbar^2}{2m c^2}\,\p_t^2\ri)\!\psi=0$} \\[8pt] \hline
		\end{tabular} }
		\caption{\footnotesize This table presents our findings within the post-Carrollian theory. To facilitate comparison, it also includes well-established formulas from the Newtonian theory, special relativity, and tachyon theory (see Figure \ref{threthe}).}
		\label{tab:example1}
	\end{center}
\end{table}

Utilizing the post-Carrollian energy-momentum relation \eqref{cem}, we applied the principles of quantum mechanics and reproduced both the Carroll–Schr\"odinger equation \eqref{cse} and the generalized Carroll–Schr\"odinger equation \eqref{inany}, obtained in \cite{Najafizadeh:2024imn}. Furthermore, we demonstrated that the post-Carrollian algebra \eqref{carrext}, rather than the Carroll algebra, governs the symmetry we are considering in this work. However, its demonstration in higher dimensions is yet unknown, as the symmetries of the generalized Carroll–Schr\"odinger equation are not determined yet. In addition, we studied the properties of an ideal gas composed of post-Carroll particles. We determined a range of thermodynamic quantities, including entropy, internal energy, specific heats, and so on \eqref{cte} within this framework.

\newpage

By analyzing the weak and static Einstein's field equations, we modeled the energy-momentum tensor using tachyon dust. Taking the ultra-relativistic limit, we derived Poisson's equation for post-Carrollian gravity \eqref{Poisson-}, along with its corresponding gravitational potential \eqref{key} as the solution. Interestingly, these results closely resemble their counterparts in the Newtonian framework. Therefore, our findings revealed that, similar to Newtonian particles, there exists an attractive gravitational force between post-Carrollian particles as well \eqref{cgf} (see also Figure \ref{3333}).

A remarkable result emerged during the derivation of the post-Carrollian gravitational field. By analyzing the geodesic equation and applying the ultra-relativistic limit, we derived a relation for the post-Carrollian gravitational field \eqref{geod5}. Surprisingly, this revealed that, unlike in the Newtonian case, the post-Carrollian gravitational field is radially outward \eqref{geod51} (Figure \ref{fig:grav_field}). This suggested that post-Carrollian particle carries a negative mass charge, while Newtonian particle possesses a positive mass charge \eqref{charge}. Furthermore, we explored the implications of this result, discussing the existence of a repulsive interaction between Newtonian and post-Carrollian particles (Figure \ref{3333}).

In this work, we have identified several physical quantities within the post-Carrollian framework that differ significantly from their Newtonian counterparts. These findings highlight the intriguing nature of post-Carrollian physics and motivate further exploration of its laws and principles. For instance, future research could explore post-Carrollian work, torque, angular momentum, and other related quantities to enhance our understanding of post-Carrollian dynamics.

The transition from special relativity to tachyon theory, achieved through the substitution $m\to im$ and the consideration of $v>c$, raises an intriguing question: Could a similar correspondence exist between Newtonian mechanics and post-Carrollian mechanics? If such a transition is possible, it suggests that a non-local field redefinition might play a crucial role. An example of this phenomenon is presented in \cite{Najafizadeh:2022vwc}, where the application of a non-local field redefinition alters the physical framework—specifically, a massive theory transitions into a massless theory through this transformation.

It is interesting to explore whether post-Carrollian particles could potentially be considered as candidates of dark matter and/or may also explain dark energy. The connection between Carroll symmetry and dark energy is discussed in \cite{deBoer:2021jej}. For instance, galaxies consist of both ordinary matter (Newtonian particles) and dark matter (post-Carrollian particles). In this scenario, the repulsive interaction between these two types of matter could explain why the dark matter halo extends significantly beyond the visible galaxy. In addition, if we assume that post-Carrollian matter exists in the universe, its repulsive interaction with Newtonian matter could account for the accelerated expansion of the universe associated with dark energy. However, further investigation is needed to fully develop and refine these ideas.

\section*{Acknowledgments}

We are grateful to Hamid Afshar, Jan de Boer, Ahmad Ghodsi, Daniel Grumiller, Mahmood Roshan, Shahin Sheikh-Jabbari, and Vahid Taghiloo for useful discussions and correspondence. We are also thankful to the {\href{https://nyuad.nyu.edu/en/academics/divisions/science/strings-conference-2025-abu-dhabi.html}{\color{black}Strings 2025 Conference (NYU Abu Dhabi, UAE, Jan 6-10, 2025)}}, where parts of this research were completed. This work is based upon research funded by Iran National Science Foundation (INSF) under project No. 4028530. The author is also supported by IPM funds.

\appendix

\section{Theories of motion: a brief review}\label{tom}

For the reader's convenience and to facilitate comparison with the main text, this appendix briefly presents the theory of special relativity in \ref{a1}, Newtonian theory in \ref{GP}, and tachyon theory in \ref{Tachyons}. This allows for a straightforward comparison between relativistic particles and tachyons, as well as between non-relativistic (Newtonian) particles and ultra-relativistic (post-Carroll) particles, as obtained in Section \ref{CP}.

\subsection{Special relativity} \label{a1}

In the context of special relativity, the total energy $\mathrm{E}$ and the momentum vector $\mathbf{P}$ of a relativistic particle, with rest mass $m$, are given by
\be 
\mathrm{E}=\ga\,mc^2\,,\qquad \mb{P}=\ga\,m\mb{v}\,, \label{p,e}
\ee 
where $\gamma$ represents the Lorentz factor
\be
\ga=\frac{1}{\sqrt{1-({v}/{c})^2}}\,, \label{gama}
\ee
and $v=|\mb{v}|=\sqrt{v_i v_i}$ is the magnitude of the velocity vector $\mb{v}$. The relativistic energy-momentum relation, which connects the total energy to the invariant mass and momentum \eqref{p,e}, is given by
\be 
\mathrm{E}^2=\mb{P}^2 c^2+m^2 c^4\,. \label{Rem}
\ee 
We note that $\mathrm{E}>|\mathbf{P}c|$ and the range of the total energy and the momentum are given by
\be 
mc^2\leqslant\mathrm{E}<\infty\,, \qquad 0\leqslant|\mathbf{P}|<\infty\,.\label{ebonded}
\ee 
This demonstrates that the total energy has a lower bound, referred to as the ``rest energy'', denoted by $\mathrm{E}_0=mc^2$, at which the particle's momentum is zero ($\mathbf{P}=0$).

By applying the quantum mechanical prescription ($\mathrm{P}_i\to -\,i\hbar \,\p_i$ and $\mathrm{E}\to i\hbar \,\p_t$) to the energy-momentum relation \eqref{Rem}, one can arrive at the Klein-Gordon equation for the scalar field $\phi$
\be 
\le(\,\Box-\frac{m^2 c^2}{\hbar^2}\,\ri)\!\phi=0\,.\label{Rqe}
\ee

Newton's second law of motion within the framework of special relativity can be derived by substituting the relativistic momentum relation \eqref{p,e} into $\mb{F}=\mathrm{d} \mb{P}/\mathrm{dt}$. It reads 
\begin{align}
	\mb{F}=\frac{\mathrm{d}}{\mathrm{dt}}(\ga\,m\mb{v})&=m\,\gamma^3\,(\mathbf{v}\cdot\mathbf{a})\,\mathbf{v}/c^2+m\,\gamma\,\mathbf{a} \nonumber\\[5pt]
	&=\ga^3\,m\,\mb{a}_{\,\shortparallel}+\ga\,m\,\mb{a}_{\!_\perp}\,, \label{rf}
\end{align}
where $\mb{a}=\frac{d\mathbf{v}}{dt}$ is the particle's acceleration. In the second line, the acceleration is separated $\mb{a}=\mb{a}_{\,\shortparallel} + \mb{a}_{\!_\perp}$ into the part parallel to the velocity and the part perpendicular to it.

\subsection{Newtonian theory} \label{GP}

Newtonian mechanics can be derived as the non-relativistic limit ($v \ll c$) of special relativity. The energy and momentum expansion of relativistic particles \eqref{p,e} can be expressed as 
\begin{align}
	\mathrm{E}~&=\qquad mc^2 \qquad+~~~\quad\frac{1}{2}\,mv^2~~~\,~+~~~\mathcal{O}(c^{-\,2})\,,\label{teee}\\[8pt]
	\mathbf{P}~&=~~~~~~~0~~\,~~~~~~+~~~~\quad m\,\mathbf{v}~~~~~~~+~~~\mathcal{O}(c^{-\,2})\,.\label{tmm}\\[-5pt]
	&~~\,~\underbrace{~~~\quad\qquad}_{\substack{\\[1pt]\text{leading terms $\equiv$}\\[2pt] \text{\!\!\!\!particle at rest}}}\,\quad \underbrace{~~~\qquad\qquad}_{\substack{\\[1pt]\text{~~~subleading terms $\equiv$}\\[2pt] \text{~~Newtonian particle}}}\nonumber
\end{align}
In these expansions, the leading terms represent the energy and momentum of Newtonian particle at rest, while the subleading terms describe moving Newtonian particle. Accordingly, ignoring the higher order terms $\mathcal{O}(c^{-\,2})$, one can define the so-called ``Newtonian energy'' $\mathrm{E}\nn$ for a free particle
\be
\mathrm{E}\nn=\mathrm{E}-\mathrm{E}_0\,,\label{sube}
\ee
which is the total energy subtracted by the particle's rest energy ($\mathrm{E}_0=mc^2$). Therefore, using the definition \eqref{sube}, one can derive the Newtonian energy $\mathrm{E}\nn$ and the Newtonian momentum $\mb{P}\!\nn$ for a non-relativistic (Newtonian) particle with mass $m$ and velocity $\mathbf{v}$ as fallows:
\be 
\mathrm{E}\nn=\frac{1}{2}\,m v^2\,, \qquad  \mb{P}\!\nn=m\,\mb{v}\,. \label{Gem}
\ee
By eliminating the velocity in the latter expressions, the Newtonian energy-momentum relation becomes
\be 
\mathrm{E}\nn=\frac{\mb{P}_{_{\!\mathrm{N}}}^2}{2m}\,.\label{Gemr}
\ee 
This result can also be derived directly from the relativistic energy-momentum relation \eqref{Rem} by considering the condition $|\mathbf{P}| \ll mc$ (which corresponds to $\mathrm{E}\nn \ll \mathrm{E}_0$) and utilizing \eqref{sube}. 

By applying the quantum mechanical prescription ($\mathrm{P}_i\to -\,i\hbar \,\p_i$ and $\mathrm{E}\to i\hbar \,\p_t$) to the Newtonian energy-momentum relation \eqref{Gemr}, one can arrive at the Schr\"odinger equation ($\nabla^2:=\p^i\p_i$) 
\be 
\le(i\hbar\p_t+\frac{\hbar^2}{2m}\nabla^2\ri)\!\psi=0\,.\label{sch}
\ee

Newton's second law of motion can be derived using the Newtonian momentum \eqref{Gem}. It becomes
\be 
\mb{F}\nn=\frac{\mathrm{d}\mb{P}{\!\nn}}{\mathrm{dt}}=m\,\frac{d\mb{v}}{dt}=m\,\mb{a}\,.\label{b4}
\ee 
Alternatively, this result can also be derived from the relativistic Newton's second law of motion \eqref{rf} by taking the non-relativistic limit.

\begin{figure}[htbp]
	\begin{center}	
		\begin{tikzpicture}
			\begin{axis}[
				xlabel={$\beta=\frac{v}{c}$},
				ylabel={$\mathscr{E}=\frac{E}{mc^2}$},
				ylabel style={at={(ticklabel* cs:0.5)}, anchor=east, rotate=-90, right=-60pt},
				xmin=0,
				xmax=3,
				ymin=0,
				ymax=3,
				legend pos=south east,
				legend style={row sep=0.3cm, font=\small},
				xtick={0,1,2},
				xticklabels={0,1,2},
				ytick={1,2,3},
				yticklabels={1,2},
				]
				
				\addplot[color=red,domain=0:5,samples=100, restrict y to domain=0:5] {1/sqrt(1-x^2)};
				\addlegendentry{$~\mathscr{E} = \frac{1}{\sqrt{1-\beta^2}}$}
				
				\addplot[color=blue,domain=0:5,samples=100, restrict y to domain=0:5] {1+x^2/2};
				\addlegendentry{$~\mathscr{E} = 1 + \frac{\beta^2}{2}$}

				\addplot[color=black,dashed,domain=0:5,samples=2, restrict y to domain=0:5] coordinates {(1, 0) (1, 5)};

				\node[fill=white, draw=red, anchor=west] at (axis cs: 1.2, 2.0) {\color{red}Relativistic particle};
				\node[fill=white, draw=blue, anchor=west] at (axis cs: 1.34, 1.5) {\color{blue}Newtonian particle};
				
				\draw[dashed, thick, red] (axis cs: 1.2, 2.0) -- (axis cs: .89, 2.3);
				\draw[dashed, thick, blue] (axis cs: 1.34, 1.5) -- (axis cs: 1.14, 1.65);
				
			\end{axis}
		\end{tikzpicture}
	\end{center}
	\caption{\footnotesize This figure shows that the velocity of a relativistic particle is bounded within the range $0 \leqslant v < c$, while the velocity of a Newtonian particle is unbounded, ranging from $0 \leqslant v < \infty$. At low velocities ($v\ll c$), the Newtonian particle curve coincides with the relativistic particle curve.}
	\label{f1}
\end{figure}
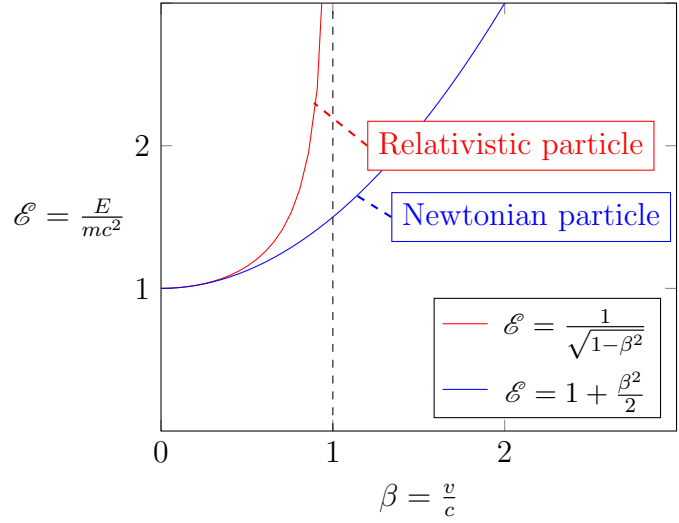

It is important to recognize that the limit $v \ll c$ serves as a technique to transition into a different physical regime. In one regime (special relativity), the velocity of a relativistic particle is limited by an upper bound and confined within the range of $0\leqslant v<c$. However, in another regime (Newtonian theory), the velocity of a Newtonian particle has no such upper bound and spans the range of $0\leqslant v <\infty$. The energy curve in terms of velocity, for both relativistic particles and Newtonian particles, is illustrated in Figure \ref{f1}. 

\subsection{Tachyon theory} \label{Tachyons}

Tachyon physics has been studied from various perspectives and motivations (see e.g. \cite{Feinberg:1967zza,PhysRevD.2.263,Olkhovsky:1971fu,Raychaudhuri:1974ew,Kirch:1975nc,Vysin:1977ue}). Tachyons are hypothetical relativistic particles that are a representation of the Poincar\'e algebra. From our conventional physics perspective, tachyons can be viewed as particles with imaginary mass that exhibit relativistic behavior by traveling at velocities greater than the speed of light ($v>c$).

To derive the tachyonic formulations, one can substitute $m \to im$ and assume that $v > c$ in the conventional relativistic equations \eqref{p,e}-\eqref{Rem}. Consequently, the Lorentz factor \eqref{gama} acquires an imaginary factor $(-\,i)$, which cancels out with the imaginary factor from the mass. Therefore, the mass parameter in tachyon relations remains a real-valued quantity, despite appearing imaginary from our perspective. According to this substitution and assumption, the energy $\mathrm{E}$ and the momentum vector $\mathbf{P}$ of a tachyon will remain real quantities, and are given by \cite{Feinberg:1967zza}
\begin{align}
	\mathrm{E} &=  \gt \, mc^2\,,\label{te}\\[5pt] 
	\mathbf{P} &= \gt \, m \mathbf{v}\,,\label{tp}
\end{align}
where $m$ is a real proper mass \cite{Recami:1984xh} and $\gt$ is the tachyon Lorentz factor defined by
\begin{align}
	\gt = \frac{1}{\sqrt{(v/c)^2-1}}\,. \label{gamma}
\end{align}
Here, $v = |\mathbf{v}| = \sqrt{v_i v_i}$ denotes the magnitude of the velocity vector $\mathbf{v}$, ranging from $c<v<\infty$.

The tachyon energy-momentum relation, which establishes a connection between the energy \eqref{te}, momentum \eqref{tp}, and invariant mass, is given by \cite{Feinberg:1967zza}
\begin{equation}
	\mathbf{P}^2 c^2 = \mathrm{E}^2 + m^2 c^4. \label{tem}
\end{equation}
As observed here, compared to the relativistic case \eqref{Rem}, the roles of energy and momentum are interchanged between relativistic particles and tachyons. Thus, while the term ``energy'' in \eqref{Rem} represents the total energy, in this case, ``momentum'' corresponds to the total momentum.

According to \eqref{tem}, one has $|\mathbf{P}c|>\mathrm{E}$ and the range of energy $\mathrm{E}$ and total momentum $\mathbf{P}$ are given by
\be 
0\leqslant\mathrm{E}<\infty \,,\qquad mc\leqslant|\mathbf{P}|<\infty\,.
\ee 
In analogy with special relativity \eqref{ebonded}, this demonstrates that the total momentum of a tachyon has a lower bound, which we call ``rest momentum'' and denote its magnitude by $|\mathbf{P}_0|=mc$. By the rest momentum, we mean that when the energy of a tachyon is zero ($\mathrm{E} = 0$), its momentum has a magnitude of $|\mathbf{P}_0| = mc$. We introduce this terminology in analogy to the concept of ``rest energy'' in special relativity. Therefore, unlike relativistic particles, tachyons have no rest energy; rather, they have rest momentum, at which their velocity is infinite (or the energy is zero). When $v\to\infty$, it follows that $\mathrm{E}=0$ and $|\mathbf{P}_0|=mc$, demonstrating that tachyons at infinite velocity carry momentum but no energy \cite{Feinberg:1967zza}. These cases represent the leading terms in the expansion of the tachyon's energy \eqref{tee} and momentum \eqref{tm}, which correspond to the energy and momentum of the magnetic Carroll particle. In other words, the magnetic Carroll particle has zero energy, $E=0$, and a rest momentum given by $\mathbf{P}=mc\,\mathbf{\hat{v}}$.

To formulate a quantum field equation within the tachyon framework, we can apply the quantum mechanical prescription ($\mathrm{P}_i\to i\hbar \,\p_i$ and $\mathrm{E}\to-\,i\hbar \,\p_t$) to the energy-momentum relation \eqref{tem}. By defining the d'Alembertian operator as $\Box:=-\,\frac{1}{c^2}\,\p_t^2+\p^i\p_i$, this procedure yields the Klein-Gordon equation of the relativistic tachyon scalar field
\be 
\le(\,\Box+\frac{m^2 c^2}{\hbar^2}\,\ri)\!\phi=0\,.\label{tqe}
\ee

By substituting the formula for tachyon momentum \eqref{tp} into the definition of Newton's second law \eqref{nsl}, we can conveniently derive the form of Newton's second law of motion for tachyons, which is
\begin{align}
	\mathbf{F}=\frac{\mathrm{d}}{\mathrm{dt}}\,(\gt\,m\mathbf{v})&=-\,m\,\gt^3\,(\mathbf{v}\cdot\mathbf{a})\,\mathbf{v}/c^2+\gt\,m\,\mathbf{a}\nonumber\\[5pt]
	&=-\,m\,\gt^3\,\mb{a}_{\,\shortparallel}+\gt\,m\,\mb{a}_{\!_\perp}\,,\label{tforc}
\end{align}
where $\mb{a}=\frac{d\mathbf{v}}{dt}$ is the tachyon's acceleration. In the second line, the tachyon's acceleration is separated $\mb{a}=\mb{a}_{\,\shortparallel} + \mb{a}_{\!_\perp}$ into the part parallel to the velocity and the part perpendicular to it.

\section{Generalized Carroll–Schr\"odinger equation}  \label{hd}

In this appendix, we reproduce the generalized Carroll–Schr\"odinger equation \cite{Najafizadeh:2024imn}, which applies to higher spacetime dimensions. In this case, the energy-momentum relation \eqref{cem} can be expressed as ($i=1,\ldots,d$)
\be 
\sqrt{\mathrm{P}^i\!\!\pc\,\mathrm{P}_{_{\!\!\mathrm{PC}}i}}=\frac{\le(\mathrm{E}\pc\ri)^2}{2mc^3}\,.\label{eman}
\ee   
Then, by keeping the energy prescription as in \eqref{pre}, it becomes clear that the momentum prescription should adopt an unconventional form. In fact, within the post-Carrollian framework, we introduce the quantum mechanical prescription as follows
\be 
\boxed{\mathrm{P}^i\!\!\pc \longrightarrow i\hbar \,\nabla^{\,i}~, \quad\qquad \mathrm{E}\pc\to-\,i\hbar \,\p_t\,,}\label{cpr}
\ee
where 
\be 
\quad \nabla^{\,i}=\frac{x^i}{x^2}\,\le(x\cdot\p+\frac{d-1}{2}\ri)\,.
\ee 	 
Taking this into consideration, one can derive the following expression
\be 
\sqrt{\nabla^{\,i}\nabla_{i}}=\frac{1}{\sqrt{x^2}}\le(x\cdot\p+\frac{d-1}{2}\ri)\equiv\nabla_x\,.\label{identi}
\ee 
As a result, by applying the prescription \eqref{cpr} into \eqref{eman} and utilizing \eqref{identi}, we can reproduce the generalized Carroll–Schr\"odinger equation \cite{Najafizadeh:2024imn}, which is
\be 
{
	\begin{aligned}
		&\le(i\hbar c\,\nabla_x+\frac{\hbar^2}{2m c^2}\,\p_t^2\ri)\psi=0\,, \label{inany}\\[10pt]
		&\text{with} \quad \nabla_x=\frac{1}{\sqrt{x^2}}\le(x\cdot\p+\frac{d-1}{2}\ri)\,.
	\end{aligned}
}
\ee
We note that in $1+1$ dimensions, where $d=1$, the operator $\nabla_x$ simplifies to the usual derivative $\p_x$, thereby reducing the equation \eqref{inany} to the Carroll–Schr\"odinger equation \eqref{cse}.

In the post-Carrollian framework, we notice that the momentum prescription indicated by \eqref{cpr} represents a deviation from the standard momentum prescription ($\mathrm{P}_i \to i\hbar\,\partial_i$). One possible explanation for this deviation can be explained as follows. When transitioning from the relativistic framework to the Newtonian framework, the momentum $\mb{P}$ experiences no shift, while the energy undergoes a shift $\mathrm{E}\nn=\mathrm{E}-mc^2$ \eqref{sube}, which is a scalar shift. However, when transitioning from the tachyon framework to the post-Carrollian framework, energy experiences no shift, but momentum undergoes a shift given by $\mathbf{P}\!\pc=\mathbf{P}-mc\,\mathbf{v}/v$ \eqref{cm}, which can be regarded as a vector shift. This distinction in the nature of the shifts could account for the modified form of the momentum prescription as given by \eqref{cpr}.

Finally, as demonstrated in \eqref{1d}, we note that the post-Carrollian momentum is always positive, leading to the selection of the prescription outlined in \eqref{pre} or \eqref{cpr}. However, in the Newtonian case, we recall that the energy is always positive \eqref{Gemr}, necessitating a reversal of the sign in the prescription, i.e. $\mathrm{E}\nn\to i\hbar \,\p_t$ and $\mathrm{P}\!\nn\to -\,i\hbar\,\p_x$.

\section{Newtonian gravity} \label{GGP1}

In this section, we will review the derivation of the Newtonian gravitational potential from Einstein's field equations in Section \ref{GGP} and the gravitational field from the geodesic equation in Section \ref{ggfg}. Our goal is to use a similar approach in Section \ref{CGFo1} to derive the post-Carrollian counterpart.

\subsection{Gravitational potential} \label{GGP}

As discussed in Section \ref{CGFo}, we can proceed from the weak and static field equations \eqref{LE2} by using an appropriate energy-momentum tensor. As well established in the literature, the weak and static field equations \eqref{LE2} can be coupled to the relativistic dust 
\be 
T^{\m\n}=\r\,U^\m U^\n\,,\label{emtr}
\ee 
where $\r>0$ represents the rest mass density and $U^\m$ is the timelike four-velocity vector, given by
\be 
U^\m=(c\,\gamma, v^i\,\gamma)\,,\label{fourv}
\ee  
with
\be 
\gamma=\frac{1}{\sqrt{1-v^2/c^2}}\,,\label{gammt}
\ee
where $v^2=v(t)^2=v^i(t)v_i(t)$ can, in principle, be time-dependent. We note that the four-velocity vector $U^\mu$ is the tangent vector at any point of world line $x^\m(\tau)$ defined as $U^\m={dx^\m}/{d\tau}$, where $x^\m = (ct,x^i)$ is the four-position vector and $\tau$ is the proper time, defined by $dt/d\tau=\gamma$. More precisely, the components of \eqref{fourv} are $U^0=\frac{dx^0}{d\tau}=c\,\frac{dt}{d\tau} = c\,\gamma$ and $U^i=\frac{dx^i}{d\tau}=\frac{dx^i}{dt}\,\frac{dt}{d\tau}= v^i\gamma$, where $v^i={dx^i}/{dt}$.

By substituting \eqref{fourv} into \eqref{emtr}, the components of the energy-momentum tensor, and its trace $T=g_{\m\n}\,T^{\m\n}\approx (\eta_{\m\n}+h_{\m\n}) T^{\m\n}$ up to zeroth order in $h$, become
\begin{align}
	&T_{00}=\r\,c^2\,\gamma^2\,\approx\,\r\,c^2\,, \label{T1}\\[8pt]
	&T_{i0}=\r\,c\,v^i\gamma^2\,\approx\,\r\,c\,v^i\,, \label{T2}\\[8pt]
	&T_{ij}=\r\,v^i v^j\,\gamma^2\,\approx\,\r\,v^i v^j\,, \label{T3}\\[8pt]
	&T=\eta_{\m\n}\,T^{\m\n}=\eta_{00}\,T^{00}+\eta_{ij}\,T^{ij}=\r\,\gamma^2\le(-\, c^2+\delta_{ij}v^i v^j\ri)=-\,\r\,c^2\,.\label{T}
\end{align}
The approximate symbols ($\approx$) in \eqref{T1}-\eqref{T3} indicate that the non-relativistic limit ($v\ll c$) or the Newtonian limit has been taken.  

By employing \eqref{T1} and \eqref{T}, the $00$-component of the weak and static field equations \eqref{LE2}, up to the order of $\r\,G\N/c^2$, reduces to
\be 
\nabla^2 h_{00}=-\,\frac{8\pi \r\,G\N}{c^2}\,.\label{h00}
\ee 
Substituting 
\be h_{00}=-\,\frac{2\phi}{c^2}\,,\label{h000}
\ee
into the equation \eqref{h00}, we recover Poisson's equation for Newtonian gravity, along with the Newtonian gravitational potential as its solution: 
\be 
\nabla^2 \phi={4\pi \r\,G\N}\,,\qquad\qquad \phi=-\,\frac{M G\N }{r}\,.\label{Poisson}
\ee
The gravitational energy potential $U$ and the gravitational force $\mathbf{F}$ then become 
\be 
U=m\,\phi=-\,\frac{m\,M\,G\N}{r}\,,\qquad\qquad \mathbf{F}=-\,\nabla U=-\,\frac{m\,M\,G\N}{r^2}\,\mathbf{\hat{r}}\,,\label{Gu}
\ee 
demonstrating an attractive gravitational force between two Newtonian particles with masses $m$ and $M$.

By applying \eqref{T2} and \eqref{T}, the $i0$-component of the weak and static field equations \eqref{LE2} contains no terms of the order $\r\,G\N/c^2$. Therefore $\nabla^2 h_{i0}=0$, demonstrating that $h_{i0}=0$ or $g_{i0}=\eta_{i0}+h_{i0}=0$.

In addition, using \eqref{T3} and \eqref{T}, the $ij$-component of the weak and static field equations \eqref{LE2}, up to the order of $\r\,G\N/c^2$, simplifies to 
\be 
\nabla^2 h_{ij}=-\,\frac{8\pi \r\,G\N}{c^2}\,\delta_{ij}\,,
\ee 
which implies that $h_{ij}=-\,\frac{2\phi}{c^2}\,\delta_{ij}$. Accordingly, recalling $h_{00}=-\,\frac{2\phi}{c^2}$, we get
\begin{align}
	g_{00}&=-\,1+h_{00}\,,\nonumber\\[5pt]
	g_{i0}&=0\,,\label{gmn}\\[5pt]
	g_{ij}&=\le(1+h_{00}\ri)\delta_{ij}\,.\nonumber
\end{align}
As a results, the full line-element that solves the linearized Einstein's equations and describes the geometry generated by a weak and static field is given by
\begin{align} 
	ds^2=g_{\m\n}dx^\m dx^\n=-\le(1+\frac{2\phi}{c^2}\ri)\,c^2 dt^2+\le(1-\,\frac{2\phi}{c^2}\ri)\,dx^i dx_i\,,
\end{align}
where $\phi$ is a solution of the Poisson's equation \eqref{Poisson}.

\subsection{Gravitational field} \label{ggfg}

Our approach to deriving the Newtonian gravitational field from the geodesic equation differs slightly from traditional approaches in the literature, as we take the non-relativistic limit at the final stage. As illustrated below, it is structured to be comparable to the derivation of the post-Carrollian gravitational field discussed in Section \ref{CGF}. The geodesic equation is given by
\be 
\frac{d^2x^{\mu}}{d\tau^2} + \Gamma^{\mu}_{\nu\rho}\frac{dx^{\nu}}{d\tau}\frac{dx^{\rho}}{d\tau} = 0\,,\label{geo11}
\ee 
or (when separating the summing indices into $0$ and $i$)
\be 
\frac{d^2x^{\mu}}{d\tau^2} 
+\Gamma^{\mu}_{ij}~\frac{dx^{i}}{d\tau}\frac{dx^{j}}{d\tau}
+2\,c\,\Gamma^{\mu}_{0i}~\frac{dt}{d\tau}\frac{dx^{i}}{d\tau}
+c^2\,\Gamma^{\mu}_{00}~\frac{dt}{d\tau}\frac{dt}{d\tau} = 0\,,\label{geo111} 
\ee
where $x^\m=(ct,x^i)$, $\tau$ is the proper time, and $\Gamma^{\mu}_{\nu\rho}$ are Christoffel symbols
\be 
\Gamma^{\mu}_{\nu\rho}=\frac{1}{2}\,g^{\m\lambda}\Big({\p_\nu} g_{\rho\lambda}+\p_\rho g_{\nu\lambda}-\p_\lambda g_{\nu\rho}\Big)\,.\label{crif1}
\ee 
By employing $dt/d\tau=\gamma$, where $\gamma$ defined in \eqref{gammt}, and applying the chain rule, we can write 
\begin{align}
	\frac{dx^{\mu}}{d\tau}&=\gamma\,\frac{dx^{\mu}}{dt}\,,\label{chain0}\\[8pt]
	\frac{d^2x^{\mu}}{d\tau^2}&=\gamma^2\,\frac{d^2x^{\mu}}{dt^2}+\frac{\gamma^4}{c^2}\,(\mathbf{v}\cdot\mathbf{a})\,\frac{dx^{\mu}}{dt}\,,\label{chain}
\end{align}
where 
\be 
\frac{d\gamma}{dt}=\frac{d\gamma}{dv^i}\,\frac{dv^i}{dt}=\frac{\gamma^3}{c^2}\,(\mathbf{v}\cdot\mathbf{a})\,,
\ee
is utilized, with $\mathbf{v} \cdot \mathbf{a}$ representing the inner product of the velocity vector $v^i=dx^i/dt$ and acceleration vector $a^i=dv^i/dt$. Accordingly, using \eqref{chain0} and \eqref{chain}, the geodesic equation \eqref{geo111} can be rewritten as
\be 
\frac{d^2x^{\mu}}{dt^2}+\frac{\gamma^2}{c^2}\,(\mathbf{v}\cdot\mathbf{a})\,\frac{dx^{\mu}}{dt} 
+\Gamma^{\mu}_{ij}~v^i\,v^j
+2\,c\,\Gamma^{\mu}_{0i}~v^i
+c^2\,\Gamma^{\mu}_{00}= 0\,.\label{geo1111} 
\ee
The Newtonian limit of the geodesic equation \eqref{geo11}/\eqref{geo1111} is acquired by three assumptions: ($i$) the gravitational field is weak, so it can be treated as a small perturbation to the flat spacetime metric $g_{\m\n}=\eta_{\m\n}+h_{\m\n}$, ($ii$) the gravitational field is static, meaning it does not vary with time $\p_0 \,g_{\m\n}=0$, ($iii$) the particles are moving slowly, with velocities much less than the speed of light $v\ll c$. 

The first assumption simplifies the Christoffel symbols to $\Gamma^{\mu}_{\nu\rho}=\frac{1}{2}\,\eta^{\m\lambda}({\p_\nu} h_{\rho\lambda}+\p_\rho h_{\nu\lambda}-\p_\lambda h_{\nu\rho})$. Then, applying the requirements $h_{i0}=0$ and $h_{ij}=\delta_{ij}h_{00}$ derived in the previous subsection, along with the second assumption $\p_0 \,h_{\m\n}=0$, the components of the Christoffel symbols become:
\begin{align}
	&\Gamma^{0}_{ij}=\Gamma^{i}_{0j}=\Gamma^{0}_{00}=0\,,\nonumber\\[3pt]
	&\Gamma^{0}_{0i}=\Gamma^{i}_{00}=-\,\frac{1}{2}\,\p_i\, h_{00}\,,\nonumber\\[3pt]
	&\Gamma^{k}_{ij}=\frac{1}{2}\le(\delta_{ik}\,\p_j+\delta_{jk}\,\p_i-\delta_{ij}\,\p_k\ri)h_{00}\,.
\end{align}
Using these components, the geodesic equation \eqref{geo1111} for $\m=0$ leads to the condition
\be 
\frac{\gamma^2}{c^2}\,(\mathbf{v}\cdot\mathbf{a})=v^i\p_i h_{00}\,. \label{cond}
\ee 
For $\m=k$, the geodesic equation \eqref{geo1111} gives
\be 
\frac{d^2x^{k}}{dt^2}+2\,\frac{\gamma^2}{c^2}\,(\mathbf{v}\cdot\mathbf{a})\,v^k 
=-\,\p_k \phi-\,\frac{v^2}{c^2}\,\p_k \phi\,,\label{ge}
\ee 
where the condition \eqref{cond} and $h_{00}=-\,\frac{2\phi}{c^2}$ have been applied. At this stage, we take the non-relativistic limit ($v \ll c$), which eliminates the second terms on both sides of \eqref{ge}, ultimately leading to
\be 
\frac{d^2x^{k}}{dt^2}=-\,\p_k \phi\,, \qquad \text{or} \qquad 
\frac{d^2\mathbf{x}}{d t^2}=-\,\mathbf{\nabla}\phi\,. \label{gf}
\ee 
By defining the Newtonian gravitational field as ${g}\N^k:={d^2{x^k}}/{d t^2}$ and substituting the Newtonian gravitational potential \eqref{Poisson}, $\phi=-\,{M G\N}/{r}$, into \eqref{gf}, we obtain the Newtonian gravitational field
\be 
\mathbf{g}\nn:=\frac{d^2\mathbf{x}}{d t^2}=-\,\mathbf{\nabla}\phi=-\,\frac{M G\N}{r^2}~\mathbf{\hat{r}}\,,\label{GG}
\ee  
indicating that the Newtonian gravitational field is radially inward, see Figure \ref{fig:grav_field}.

An important result here is that the term containing $\mathbf{v}\cdot\mathbf{a}$ in \eqref{ge} ignores in the non-relativistic limit. Consequently, the gravitational field \eqref{GG} remains independent of the direction of the test particle's velocity and acceleration, a property that plays a crucial role in the post-Carrollian gravitational field.

\bibliographystyle{hephys}
\small\bibliography{references}

\end{document}